# The Hamburg Meteorite Fall: Fireball trajectory, orbit and dynamics

P.G. Brown[1,2*], D. Vida[3], D.E. Moser[4], M. Granvik[5,6], W.J. Koshak[7], D. Chu[8], J. Steckloff[9,10], A. Licata[11], S. Hariri[12], J. Mason[13], M. Mazur[3], W. Cooke[14], and Z. Krzeminski[1]

***Corresponding author email: pbrown@uwo.ca**

**ORCID ID:** https://orcid.org/0000-0001-6130-7039

[1]Department of Physics and Astronomy, University of Western Ontario, London, Ontario, N6A 3K7, Canada

[2]Centre for Planetary Science and Exploration, University of Western Ontario, London, Ontario, N6A 5B7, Canada

[3]Department of Earth Sciences, University of Western Ontario, London, Ontario, N6A 3K7, Canada (

[4]Jacobs Space Exploration Group, EV44/Meteoroid Environment Office, NASA Marshall Space Flight Center, Huntsville, AL 35812 USA

[5]Department of Physics, P.O. Box 64, 00014 University of Helsinki, Finland

[6] Division of Space Technology, Luleå University of Technology, Kiruna, Box 848, S-98128, Sweden
[7]NASA Marshall Space Flight Center, ST11, Robert Cramer Research Hall, 320 Sparkman Drive, Huntsville, AL 35805, USA

[8]Chesapeake Aerospace LLC, Grasonville, MD 21638, USA

[9]Planetary Science Institute, Tucson, AZ, USA

[10]Department of Aerospace Engineering and Engineering Mechanics, University of Texas at Austin, Austin, TX, USA

[11]Farmington Community Stargazers, Farmington Hills, MI, USA

[12]Department of Physics and Astronomy, Eastern Michigan University, Ypsilanti, MI, USA

[13]Orchard Ridge Campus, Oakland Community College, Farmington Hills, MI, USA

[14]NASA Meteoroid Environment Office, Marshall Space Flight Center, Huntsville, Alabama 35812, USA

## Abstract

The Hamburg (H4) meteorite fell on January 17, 2018 at 01:08 UT approximately 10km North of Ann Arbor, Michigan. More than two dozen fragments totaling under one kilogram were recovered, primarily from frozen lake surfaces. The fireball initial velocity was 15.83 ± 0.05 km/s, based on four independent records showing the fireball above 50 km altitude. The radiant had a zenith angle of 66.14˚± 0.29˚and an azimuth of 121.56˚± 1.2˚. The resulting low inclination (<1°) Apollo-type orbit has a large aphelion distance and Tisserand value relative to Jupiter ($T_j$) of ~3. Two major flares dominant the energy deposition profile, centred at 24.1 and 21.7 km altitude respectively under dynamic pressures of 5-7 MPa. The Geostationary Lightning Mapper (GLM) on the Geostationary Operational Environmental Satellite - 16 also detected the two main flares and their relative timing and peak flux agree with the video-derived brightness profile. Our preferred total energy for the Hamburg fireball is 2 – 7 T TNT ($8.4\text{-}28\times10^9$ J), which corresponds to a likely initial mass in the range of 60 – 225 kg or diameter between 0.3 – 0.5 m. Based on the model of Granvik et al (2018), the meteorite originated in an escape route from the mid-outer asteroid belt. Hamburg is the 14$^{th}$ known H-chondrite with an instrumentally-derived pre-atmospheric orbit, half of which have small (<5°) inclinations making connection with (6) Hebe problematic. A definitive parent body consistent with all 14 known H-chondrite orbits remains elusive.

## Introduction

Measuring the pre-impact orbits for meteorites provides a unique linkage between laboratory-based meteorite studies and asteroid science. Meteorite orbits immediately prior to Earth impact represent the final stage in a long stochastic process of orbital migration of meteorites from the main asteroid belt to Earth (Vokrouhicky et al., 2000). With sufficient numbers of meteorite orbits, statistical inferences as to the origin of particular meteorite group source regions in the main-belt may be made (Granvik and Brown, 2018) which ultimately may provide unique constraints as to the original parent bodies for some groups of meteorites.

As of late-2018, 27 meteorite orbits have been published (Granvik and Brown, 2018; Borovička et al., 2015; Devillepoix et al., 2018) with at least seven more having sufficient data for orbit determination but not yet published. Among these are thirteen H-chondrites with measured orbits, by far the largest of any meteorite group.

The main-belt source region of the H chondrites remains unclear. Gaffey and Gilbert (1998) proposed the large main-belt asteroid (6) Hebe to be the primary parent body for the H-chondrites, based on its similar reflectance spectra to H-chondrites, size and location near the 3:1 (Jovian) mean-motion resonance (MMR) and the $\upsilon_6$ secular resonance, both major escape routes from the main asteroid belt. More recently, other main-belt asteroids with reflectance spectra consistent with H chondrites have also been found (Vernazza et al., 2014) throughout the main-belt. Fieber-Beyer and Gaffey (2014) found several small, H-chondrite-like asteroids proximal to (6) Hebe and suggested a family linkage. Most recently, however, Marsset et al. (2017) used high resolution imagery of (6) Hebe to rule out an impact basin large enough to be consistent with the total volume of these nearby smaller H-chondrite-like asteroids. Finally, NEAs found to have compatible H-chondrite reflectance spectra also appear to be most likely delivered from the 3:1 MMR, but with some contribution from the 5:2 MMR and $\upsilon_6$ possible (Binzel et al., 2015). Taken together, current evidence suggests escape routes for H-chondrites ranging in distance from the 3:1 to as far as the 5:2 MMR resonance are all plausible candidates; the role of (6) Hebe in the delivery picture remains unclear.

Here we describe instrumental records of the fall of the Hamburg (H4) meteorite (Heck and Greer, 2018). These data include direct camera recordings of the fireball, brightness measurements of the fireball recorded by the Geostationary Lightning Mapper (GLM) on the GOES-16 satellite, Doppler weather data recording the atmospheric meteorite debris plume as well as meteorite recovery locations. From these data we have estimated the pre-atmospheric orbit and initial mass of the Hamburg meteorite and use this information to estimate the probability that Hamburg is derived from (6) Hebe in addition to its relationship to escape routes common to other H-chondrites.

## General circumstances of the meteorite fall and overview of instrumental records

The Hamburg fireball occurred at 01:08:29 UTC (20:08:29 EST) on Jan 17, 2018 (Jan 16 local time) over Southeastern Michigan, USA. The fireball was widely seen by numerous eyewitnesses, with more than 650 of submitting fireball reports to the American Meteor Society (Perlerin, 2018) from seven US states and the Canadian province of Ontario. Sonic booms from the fireball were also reported by many eyewitnesses and seismo-acoustic coupling of this shock was detected at a nearby infrasonic and seismic stations (Hedlin et al., 2018) while the infrasonic wave was also detected at the ELFO infrasound array (Silber and Brown, 2014) in Ontario. Additionally, Doppler weather radar detected the falling debris plume of meteorites shortly after the fireball. Cloud conditions at the time were unfavorable – most of Southern Ontario and large portions of Michigan and Ohio were nearly overcast making direct visibility of the fireball challenging. Figure 1 shows the cloud conditions in the region at the time of the fireball.

Despite the poor weather, many casual video recordings of the meteor were secured. In total we geolocated 27 unique videos showing either the direct fireball or its indirect scattered light. The all-sky cameras of the Southern Ontario Meteor Network (Brown et al., 2010) and NASA All Sky Fireball Network (Cooke and Moser, 2012) were largely overcast at the time, though many stations showed two distinct flashes as the fireball illuminated clouds at ranges in excess of 400 km, providing timing checks on the two major flares. One NASA all sky camera in Oberlin, OH (Figure 2a) had partial clear sky and recorded much of the fireball through thin, high clouds.

From among these 27 casual video recordings, four were selected which showed a direct view of the fireball and were in fixed positions allowing direct astrometric calibration. In all four cases we were able to later obtain nighttime stellar calibration data for the same systems. All cameras had generally unchanged pointing since the time of the fireball permitting plate calibrations; small changes in some calibration images were corrected manually. Two additional records were selected for use in relative photometry measurements. These two were selected because one contained a local light source which we used to remove the effects of the cameras automatic gain control on the photometry from the relative brightness of scattered light on the ground while the other one (the Oberlin all-sky video) had an unsaturated direct view of the fireball for the early to middle portion of the trajectory. Table 1 lists the details of these five casual video records and the Oberlin all-sky camera data.

The most complete record of the fireball among these calibrated videos were from Defiance, OH, Chicago, IL, Brant, ON and Madison, WI. That the best recordings are from very distant cameras reflects the heavy cloud cover in the vicinity of the fireball trajectory, which greatly reduced visibility of the fireball for all cameras with ranges under 100 km. The Oberlin, OH NASA all-sky camera detected the fireball behind thin, high cloud in the early portion of the trajectory and had a substantially worse pixel scale than the other cameras; it was not used in the final astrometric solution. Figure 2 displays individual frames showing the fireball from early portions of Oberlin, Defiance, Chicago and Madison where the images are unsaturated. Two bright flares dominate the late stages of flight for all cameras, with the fireball remaining luminous to a height of 19.7 km.

## Fireball Trajectory and Orbit

All cameras had positional calibrations using background stars for each of the four fixed cameras, generally taken within a few days to at most a few weeks of the event. In most cases the cameras remained fixed and no noticeable change in orientation was apparent using nearby objects as a guide. Astrometric plates were computed using two 3$^{rd}$ order polynomial fits as described in Weryk and Brown (2012). Although the individual frame picks were often uncertain due to clouds or blooming, the stations had good intersection geometry and small pixel scales. As a result, we found the trajectory orientation to be quite robust (less than a degree of variance) under inclusion of different combinations of stations and points. Figure 3 shows the camera locations relative to the fireball path and meteorite strewn-field. The details of the calibration for each of these four sites, including plots of plate residuals are given in appendix A. The appendix also contains plots and tables related to the final trajectory and orbit as well as all individual astrometric picks for all four stations. Note that for the Madison, WI site, a radial plate fit (Borovička, 2014) was found to produce slightly better agreement with the trajectory as compared to other sites and this was used for the final trajectory solution.

Figure 4 shows the resulting lateral deviations (positive indicating a sightline passed above the final trajectory) from a straight-line fit. We show all four stations where measurements were made. All station weights were equal. The trajectory was computed using a line-of-sight (LoS) trajectory optimization routine similar to Borovička (1990). As described in Vida et al., (2019) this technique uses a Monte Carlo formalization of the LoS solver, estimating the mathematical fit uncertainty based on the variance of measurements from each station and implementing Earth rotation at all points. This produces a lower bound to the total uncertainty in the trajectory measurement. The best estimate for the orbital and radiant uncertainties are based on the variation in the trajectory found by iteratively removing outlying points from various stations and noting the maximal change to the radiant and velocity. Finally, an approximate upper bound to the speed uncertainty by using the standard deviation of the individual station speed measurements (see appendix A, Table A3 for details). Table 2 summarizes the final trajectory solution.

The estimation of the fireball orbit is critically dependent on the initial velocity; that is the velocity before measureable deceleration takes place due to atmospheric drag. This height depends on the camera system and its resolution and the geometry and number of cameras relative to the trajectory. It also depends on the meteoroid mass, speed and entry angle (e.g. Vida et al., 2018). This deceleration height varies from over 100 km altitude for small, slow, shallow-entry meteoroids, to very low heights for larger (decimeter to meter-sized), steeper events. We found that noticeable deceleration did not become apparent in the solutions for the steeply-entering Hamburg fireball until below 50 km altitude. This is consistent with our entry modelling (see discussion) which predicts total deceleration was less than 0.1 km/s above 50 km height. The four astrometrically calibrated cameras had average ECF speed estimates between 15.6 – 15.9 km/s based on the trajectory visible from each station above 50 km height, with an average speed of 15.78 ± 0.14 km/s. The uncertainty here represents the spread in the independent speeds across all four cameras and does not account for any other systematic effects. However, that four independently calibrated cameras all produce similar initial average speeds provides confidence this is a physically reasonable range for the initial speed. The final best estimate of the initial velocity based on simultaneous fitting of the length versus time for all cameras including timing offsets using the method described in Vida et al (2019) is 15.83± 0.05 km/s. The uncertainty here represents the spread in speeds for different solutions where outlying points are removed, capturing an estimate of the systematic errors present. The resulting orbit, computed following the procedure in Ceplecha (1987), is given in Table 3.

The computed orbit has an aphelion close to Jupiter. The Tisserand value with respect to Jupiter is 2.99±0.003. Such an orbit is borderline between asteroidal and Jupiter-family comet-type orbits and would appear to be odd for an H-chondrite. However, of the thirteen previously published H-chondrite orbits, two have similarly low $T_j$ of ~3 values (Ejby and Kosice) and two others have $T_j<3.1$ (Benesov and Murrili) (Granvik and Brown, 2018).

## Fireball Lightcurve

### Video Data

Reconstruction of the fireball lightcurve from video observations was challenging due to cloud conditions and saturation. Two strong flares near the end of the trajectory were visible on most videos and are localized in time to within one video frame in all videos. We were able to use the timing of these two flares to compute the relative timing between each video. Our reconstruction of the fireball lightcurve from direct and indirect (scattered) intensity is shown in Fig. 5. Our approach was to use the relative change in brightness (either direct or scattered from surfaces in the video) from the most reliable video segments and assume that all offsets between the videos are multiplicative so that the slopes are matched. This is broadly similar to the approach used for the Chelyabinsk (Brown et al., 2013), Kosice (Borovička et al., 2013a) and Jesenice (Spurny et al., 2010) meteorite-producing fireballs.

The earliest portion of the fireball was directly visible from Madison, WI. Several stars visible in the same frames at almost the same altitude as the fireball were used to calibrate the absolute brightness of the rising portion of the fireball using standard meteor aperture photometry techniques (e.g. Weryk and Brown, 2012) from this distant station for the first second of visibility of the fireball. After this stage the fireball saturates the video and is also very low from the Madison station. The unsaturated portions of the direct fireball image from Oberlin and Defiance are used from t=1 sec to t=2.5 sec after which time all stations become saturated. From t=2.5 sec to t=4.2 sec, indirect scattering from two regions of snow covered roof surfaces as visible from the nearby Ypsilanti, MI video was used following the approach of Brown et al (2013) for Chelyabinsk exploiting a directly visible lightsource to calibrate changes in the automatic gain control for the camera (see Figure 6). The latter portion of the fireball from Defiance was also unsaturated and scaled to match the Ypsilanti second flare to produce the end portion of the lightcurve to t=4.5sec. Note that the timing after the second flare is uncertain due to changes in the frame rate of the Defiance camera (as described in the appendix). The value shown here is a lower limit.

Matching the lightcurve peaks to the spatial trajectory solution we find that the first flare occurred at a height of 24.1 km at a time of 01h 08m 33.44sec UT and the second at a height of 21.7 km at 01h 08m 33.64sec UT. Hedlin et al (2018) estimated a best fit flare height between 22-24 km based on hypocenter solutions from acoustic arrivals detected at nearby seismic/infrasound stations, agreeing with our result.

**Optical Energy Estimate from the Geostationary Lightning Mapper (GLM)**

The Geostationary Operational Environmental Satellite - 16 (GOES-16) weather satellite has on-board the world's first Geostationary Lightning Mapper (GLM; Goodman et al. 2013), which continuously monitors total lightning (i.e., cloud-to-ground flashes and cloud flashes) across a large region of the Western hemisphere. The GOES-16 GLM has been under calibration/validation evaluation for over a year during its Post Launch Product Test (PLPT) phase and has reached the "Provisional Validation Level" at the time of this writing. The PLPT evaluation ensures that optimal products are available for both the operational forecasting and broader scientific research communities.

Interestingly, the GLM detected the Hamburg bolide, and therefore provides a unique way to investigate its optical energy characteristics and compare with the video lightcurve. This is not the first bolide to be detected and studied using GLM data; see for example the bolides investigated by Jenniskens et al. (2018).

As described in Goodman et al. (2013), GLM is a high-speed nadir-staring event detector that operates in the near infrared. The narrow band (1 nm) interference filter is centered near a prominent oxygen emission triplet in the lightning spectrum at 777.4 nm. This is the same band employed in the earlier low-Earth-orbit Lightning Imaging Sensor (LIS; Christian et al., 1999) aboard the Tropical Rainfall Measuring Mission (TRMM) satellite. As with LIS, the lens/filter system focuses the input optical radiation onto a high-speed Charge Coupled Device (CCD) focal plane, and signals are read out in parallel into Real-Time Event Processors (RTEPs) for pixel-level *event* detection and data compression. The GLM employs a 1372 × 1300 pixel CCD focal plane array with a pixel footprint resolution of about 8 km (at nadir) to about 14 km (at the edge of the field-of-view). The frame time of the GLM is ~ 2 ms. Similar to LIS, several techniques are used for detecting lightning both at day and night. Daytime detection is more challenging because the solar lit thundercloud tops are typically far brighter than the diffuse multiple-scattered cloud-top lightning optical emissions.

Adjacent (side-to-side, diagonal) pixel-level optical events in one GLM frame define an optical *group*, and optical groups are combined with specific distance/time constraints to define a lightning *flash*; see Mach et al. (2007) for more details, including information on how the fundamental optical event data is clustered and filtered to create the Level 2 GLM group and flash data products. The GLM flash detection efficiency (probability of detection) instrument requirement, presently being validated in the PLPT phase, is 70% with a 5% false (i.e., non-lightning) alarm rate. These values are average criteria over the field-of-view and over a 24 hr period. The GLM location accuracy at the sub-satellite point and timing accuracy, also both being validated in the PLPT phase, are to be within 5 km and 1 ms, respectively.

Fig. 7 provides a plot of the GLM-detected optical groups produced by the Hamburg bolide. The "stair-step" appearance is an artifact of the Ground Segment processing algorithm for generating the Product Distribution and Access (PDA) operational data feed during the GLM Provisional Validation level period. The instrument optical energy granularity is in reality about a factor of 100 better than the ~1.5 fJ granularity shown in Fig. 7. Nonetheless, the 1.5 fJ granularity is adequate for the energy estimates for the Hamburg fireball. Note that there is no signal during the "dead interval" from about 4.03 s to 4.14 s in the plot when the fireball emission fell below the detection threshold of the instrument.

All of the optical groups in the bolide were single event groups, except during a period from about 4.14 s to 4.18 s when all the groups had 2 events each. This is when the bolide emission crossed a pixel boundary thereby triggering the two adjacent pixels. In turn, this splits the total group energy between the pixels, and also about doubles the total solid angle of the bolide source emission.

Taking into account the GLM entrance pupil area, the pixel solid angle, and the filter bandwidth, the group energy plot can be converted into a spectral energy density (SED) plot as shown in Fig. 7 (red plot). The spectral energy density is simply a frame-time integrated radiance in units of $\mu Jm^{-2}sr^{-1}nm^{-1}$.

Converting the spectral energy density into an equivalent absolute magnitude to compare to the video lightcurve is problematic, given the narrow-bandwidth of the GLM detector. This issue is discussed and analysed extensively in Jenniskens et al (2018). Under the assumptions adopted in the Jenniskens et al. (2018) analysis of GLM bolide data and accepting their result that the limiting sensitivity for GLM is near peak visual absolute magnitude -14, we can use the spectral energy density and (assuming the floor near an SED of 15 $\mu J\ m^{-2} ster^{-1} nm^{-1}$ is this background) to convert to absolute magnitude and compare to the video lightcurve. The resulting comparison near the peak is shown in Fig 8. The agreement in relative timing and amplitude for the two flares between the video and GLM lightcurves is very good, providing some confidence in the overall shape and peak values found with the video data alone.

## Meteorite Darkflight Modelling and Comparison to Fall Ellipse

The Hamburg meteorite strewn field lay across a chain of lakes in southern Livingston County, Michigan. These lakes formed at the margin of the Saginaw and Huron lobes of the Laurentide Ice Sheet toward the end of the Wisconsianan Glacial Episode (~14 kya). The land between these lakes is heavily populated and largely privately-owned.  In addition, there is a considerable area of marshes and wetlands in the strewn field, making discovery/recovery of meteorites difficult. Thus, nearly all recovered meteorites were found on top of the lakes. At the time of the fall, these lakes had a thick ice cover (>10 cm) coated with a fresh snowfall.  For the three days following the fall, no further snow fell, and air temperatures remained below freezing. Meteorites that landed on the lakes were presented as black rocks against a flat, white background. Our group searched Bass Lake, Strawberry Lake, and Hamburg Lake for meteorites. We recovered six samples: three from Strawberry Lake, two on Hamburg Lake and one on Bass Lake. The Bass Lake find consisted of many small fragments totaling less than a gram, all within a ~2 m radius.

Four days after the fall, the air temperature warmed sufficiently to melt the ice surface, complicating further searches of the lakes. We obtained permission from Hamburg Township to search Manley Bennett Park for meteorites, which shifted our search to land, where we recovered a small meteorite. Additionally, we constructed magnetic sweepers to collect fine magnetic materials from the land surface. These collected materials were heavily contaminated with hematite and magnetite delivered by the Laurentide Ice Sheet. We manually identified and sorted these grains under a microscope, which ultimately revealed two clusterings of meteoritic fragments on the land surface, adding an additional two samples to our total.

To further complete our mapping of the strewn field, we searched local news reports and social media reports for people that found meteorites. We contacted these people to compile the mass and location of their meteorites. We ensured that these reported meteorites had been confirmed as meteoritic by a trained geologist or meteorite hunter, and only include these verified meteorites in our compiled list. Not all contacted persons responded to our enquiry. Nevertheless, this added an additional four new, unreported samples to the total (see Table 4 for masses and locations of all recovered meteorites). Finally, we include in our study the locations of recovered meteorites reported in Heck and Greer (2018).

Figure 9 shows the location of recovered meteorites in the fall area near Hamburg, MI. together with the ground projection of the fireball path.

Figure 10 shows the upper winds extracted from radiosonde data from White Lake, MI (less than 50 km range from the fireball endpoint) at 0 and 12 UT, the closest radiosonde measurements to the time of the fall at 01:08 UT. Note the large change in wind speed and directions between 5 and 10 km altitude between these times; we do not know how quickly this change occurred. We found linearly interpolating the winds to 1 UT produced darkflight results with a fall line parallel to the fireball trajectory but significantly (~1 km or more) North of Bass and Strawberry Lake, where most finds were located. In contrast, we found that interpolating the winds to 3 AM produced darkflight locations crossing Strawberry and Bass Lake. As we do not know when the wind shifted, this better agreement between darkflight model results and finds using a 3 AM interpolation is adopted as the more likely wind field, in the absence of finer temporal resolution wind measurements.

To further check that our chosen fireball terminal point is consistent with recovered fragments and attempt to estimate release altitudes for recovered fragments, we use the darkflight model of Ceplecha (1987) which accounts for atmospheric drag, winds and Earth's rotation. This is used together with the modifications discussed in Brown et al. (2011) which integrates a Monte Carlo approach allowing for a distribution of fragment shapes and velocity perturbations due to fragmentation to estimate the spread in landing locations. In these model runs, individual spherical fragments with masses from 1kg - 1g, bracketing the mass range of recoveries [100g, 0.2g], are released at the endpoint (19.7 km) and at the height of the two flares (24.1 km and 21.7 km). In all simulations, fragments are assumed to have velocities of 3 km/s at the point of release, consistent with other fireball observations indicating that this is the velocity at which luminous flight ceases (Ceplecha et al 1998) and are assumed to have a bulk density of 3400 kg $m^{-3}$ appropriate to the average for H chondrites. Note that if the speed at release is higher, ablation will tend to reduce the size of the fragment, but since the trajectory is very steep the fall point moves only slightly to the West. The nominal prediction for the Hamburg strewnfield from these simulations independent of release height produces a fall zone oriented roughly parallel to the direction of the trajectory, but offset slightly to the East and North due to the prevailing winds.

The darkflight line produced by this mass range of fragments falls directly across Bass Lake and Strawberry Lake where most meteorite recoveries were reported as shown in Figure 11, consistent with our combination of terminal fireball location and wind field. Note that changing the shape factor from that of a sphere to that of a brick (following Halliday et al., 1984) moves fragments to the East while smaller drag coefficient would move fragments to the West along this line. Given the location and masses of recoveries (Fig 9) in comparison to predictions by release altitude from the fireball, most recovered fragments are consistent with having originated from near the terminal portion of the fireball path, with some locations showing intermixing of various masses. This could indicate that some of the fragments may have undergone further fragmentation during darkflight. The finds of tens of gram sized fragments on Bass Lake are several kilometerss west of our prediction from the endpoint release. This could also indicate that our trajectory is shifted several hundred meters to the west, that these fragments reached sub-luminous flight further west than our measured end point or that the fragments have more streamlined shapes (and hence experience lower drag) than our assumptions. It is also possible that the windfield differs from our adopted model or that the ground scatter is produced by lift forces in flight during the process of fragmentation which have shifted the location of these fragments from a purely drag solution.

To estimate the expected ground scatter of individual fragments released at a given height we introduce random cross-trajectory speeds consistent with the observations reported by Borovička and Kalenda (2003) for the Morávka fireball. They measured velocities perpendicular to the main fireball trajectory averaging ~50 m/s (but extending up to 300 m/s) for individual fragments. We have used the same three release heights as shown in Fig. 11, but add to each fragment's velocity vector a randomly oriented velocity perturbation of up to 50 m/s. Our overall procedure follows a similar methodology previously applied to the Grimsby fireball (Brown et al., 2011). Fig. 12 shows the resulting spread in fragment fall locations only for terminal altitude of release. The spreads are very similar at the height of the two flares, but are omitted from this figure for clarity. The expected spreads for 10g and 100g masses may partially explain the large mass intermixing on the lakes, though the westerly finds on Bass Lake remain puzzling. In particular, the recovery of 1-2g fragments is very difficult to reconcile with the darkflight model from any release altitude; these we interpret as likely late released fragments from larger (tens of gram pieces) which fragmented during darkflight just prior to ground contact, an effect noticed in other large strewnfields (eg. Bruderheim as discussed in Folinsbee and Bayrock (1961)).

The meteorite debris "curtain" was directly observed falling to the ground by the Doppler weather radar station KDTX in Detroit, MI, located only 40 km ground range from the fall ellipse. Meteorite debris plumes have been regularly observed by the US NEXRAD Doppler radar systems (Fries and Fries, 2010), with more than a dozen meteorite falls in the US and Canada having probable NEXRAD signatures (Fries, 2018). The NEXRAD systems consist of WSR-88D Doppler weather radars operating at a wavelength of 10 cm with peak power of 750 kW and a beam width of approximately 1 degree to the 3 dB points (Crum and Alberty, 1993).

Spatial comparisons are usually made between darkflight model predictions of fall locations and meteorite recoveries, but with Doppler radar signatures both spatial and temporal constraints can be introduced to refine release heights of fragments. Such a matching procedure has previously been performed for the Grimsby meteorite fall (Brown et al., 2011).

Figure 13 shows five KDTX sweeps in temporal order starting at 01:12:35 UT and ending at 01:19:58 UT over the fall area. These are the first sweeps which show any reflectivity signal in the area. An earlier sweep covering 1km altitude occurring ~200sec after the fireball shows no detectable signal. A signal detection during this earlier sweep would correspond to multi-hundreds of gram to kilo-sized fragments released in the final 10 km height interval of the fireball luminous trajectory. This non-detection suggests few such sized fragments were present.

To associate these radar returns with probable release altitudes, we examined a range of fragment masses released at each of the flare heights and the end height to find the best match in time and space with the recorded doppler signature. Fig. 13 shows our best matches color coded to the release points. In general we are able to match the timing to within a few tens of seconds of the radar returns beginning to the north and west with masses of tens of grams to 100g from fragments released at the end point. As time progresses, smaller fragments from the end point / final flares are reasonable matches in time to the radar returns, though the model fall locations move progressively farther north than the Doppler returns. This difference is comparable to our predicted dispersion based on 50 m/s standard deviation lateral velocity spreads (see Fig 12). While the overall match is quite good, there is a northerly skew. This may reflect fragment shapes which are not spherical (and which are therefore blown more Southward by the wind) or could be due to the real ground path being several hundred meters further south than our estimate or variability in the winds.

In summary, the darkflight model and observations are consistent with a scenario whereby 100g to tens of gram masses land furthest north/west and arrive first (and are detected in the initial Doppler radar sweeps) with smaller fragments landing progressively later to the east. All of these originate from the very end portion of the trail; release heights of 27-30 km are entirely incompatible with either recovered fall locations or radar sweep timing of any of the fragments.

The mixing of recovered fragments with more than an order of magnitude differences in mass in the same locations is an expected feature of such a steep trajectory and has been noted with other steep entries, such as Kosice (Toth et al., 2015). Small differences in the drag or lift coefficient, late fragmentation during darkflight together with lateral fragmentation forces may cause such mixing

## Discussion

### Ablation entry model and initial mass/energy

We attempt to estimate the initial mass and approximate fragmentation behavior of the Hamburg fireball by modelling its dynamics and lightcurve and comparing to observations. We focus on a very global match to the overall energetics rather than trying to reconstruct highly detailed fragmentation history as has been done for other meteorite-producing fireballs (eg. Kosice; Borovička et al (2013)) as we have only coarse dynamic data from videos. In particular, we assume for simplicity that fragmentation points produce light dominated by small particles and not from large fragments.

Specifically, using the estimated initial estimated speed as discussed earlier, we employ the FM model of Ceplecha and ReVelle (2005) to match the lightcurve from Fig. 5 starting at the maximum observed height of the fireball (83 km). This is equivalent to trying to reproduce the energy deposition profile; without detailed matching of the dynamics this is not a unique solution, but should provide reasonable estimates for the total mass.

The FM model is a numerical implementation of the standard differential single-body equations of meteor flight (Ceplecha et al., 1998), taking into account explicit fragmentation at discrete points into either dust or ponderable fragments. It permits changes in the leading fragment shape-density coefficient, $K = \Gamma A \rho_m^{-2/3}$ where $\Gamma$ is the drag coefficient, A is the shape factor and $\rho_m$ is the bulk density of the meteoroid. It also allows for a variable luminous efficiency ($\tau$, where the fireball brightness in absolute magnitude units is given by $M = -0.4 (\log(\tau dE_k/dt) - 3.17)$ and $E_k$ is the instantaneous meteoroid kinetic energy in MKS units assuming a 0 magnitude meteor emits 1500W. We use a fixed intrinsic ablation coefficient of 0.004 $s^2/km^2$ and fixed shape-density coefficient (K) of 0.0046 (MKS) following the approach of Ceplecha and ReVelle (2005) and Borovička et al. (2013a). We do this to restrict the number of free parameters, recognizing that our data with few constraints would not produce meaningful fits if varying these factors. The methodology has been validated through matches with the observed flight characteristics of many well recorded fireballs, several of which have produced meteorites (Ceplecha and ReVelle, 2005).

We use the original luminous efficiency adopted in Ceplecha and ReVelle (2005), which corresponds to values from 0.3% - 1.5% over the full trajectory for Hamburg. Note that Borovička et al. (2013a) have argued that the luminous efficiency relation proposed in Revelle and Ceplecha (2001) is a better match to several recent meteorite-producing fireballs. Those values for the luminous efficiency are 3-4 times larger than the luminous efficiency used in Ceplecha and ReVelle (2005). Hence our estimated mass from fitting the lightcurve alone may be considered an upper limit, with masses as low as ~60kg possible based purely on the lightcurve fit for the highest luminous efficiency values (~5%) proposed by Borovička et al. (2013a) .

As much of the videos showing the mid trajectory are saturated we have no precise dynamic data over this interval and restrict our initial model constraint to matching the lightcurve. Figure 14 shows our best fit model lightcurve match to the observations., Here the match was done entirely by trial and error and is representative and not necessarily unique. The estimated total mass for this fit corresponds to 225 kg and the end mass to just over 1 kg. The fits are both reasonable matches to observation.

The model was able to fit the lightcurve assuming no fragmentation until 68 km height at which point a small amount of the total mass (<1%) is released as fine grains to match a small increase in the slope of the lightcurve at this height. This earliest possible fragmentation/disintegration point corresponds to a ram pressure of 27 kPa and is a similarly low value found to several other meteorite producing fireballs (eg. Kosice at 90 kPa; Borovička et al (2013a)). A second slightly more significant jump in the lightcurve near 48 km corresponds to ~1% mass loss as minor fragments/dust under a ram pressure of 0.3 MPa. However, both of these early features are marginal. While their inclusion improves the model fit to the observed lightcurve these features are also near the limit of our expected uncertainty in early lightcurve reconstruction.

However, the first major flare centred at 24.1 km altitude is well defined and begins at 26.5 km under a dynamic pressure of just over 5 MPa while the second centred at 21.7 km begins at 22.6 km at with over 7 MPa of dynamic pressure. These are the most energetic fragmentation events and reflect comparatively high dynamic pressures, similar to the main flare for the Benesov fireball (Borovička and Spurny, 1996). The model for the first flare corresponds to a loss of almost 50% of the total remaining mass at this height being consumed to produce dust/small fragments, while the second starting at 22.6 km represents a loss of >90% of the remaining mass at this height to small fragments. This leaves only ~1 kg in the main fragment after the last flare. This does suggest that one or more kilo-sized fragments may have survived after the flares and would have experienced peak dynamic pressures of ~7 MPa, similar to the compressive strengths of the largest fragments which survived other meteorite-producing fireballs (Popova et al., 2011) such as Moravka (Borovička and Kalenda, 2003) and Kosice (Borovička et al 2013a). The dynamic pressure at the flares is also similar to the dynamic pressure at the point of catastrophic disruption for Chelyabinsk, which occurred between 1-5 MPa (Borovička et al., 2013b).

The major flares suggest that most of the initial mass survived to under 30 km altitude before being consumed in the two rapid fragmentation events which likely produced most of the recovered fragments. These continued to ablate briefly before reaching sub-luminous speeds just below the flare altitudes, broadly consistent with the darkflight models.

The main conclusion from this model comparison is that we are able to explain the lightcurve by having the vast majority of the mass of the Hamburg meteoroid ablate to small particles (or dust), with the meteorites reaching the ground as ponderable fragments representing a small fraction (of order only a few percent) of the initial mass.

Our model mass estimate corresponds to a total energy for the Hamburg fireball of ~7 T TNT or $2.8\times10^{10}$ J) explosive equivalent. This can be compared to the work of Hedlin et al. (2018) who used acoustic periods measured at infrasonic stations near the fireball to estimate yield, a technique commonly applied to bolides (Ens et al., 2012). They estimated a nominal yield for Hamburg to be ~2 T TNT equivalent, with uncertainty bounds ranging from 1-8 tons TNT. Our estimate overlaps at the high end with that of Hedlin et al. (2018). We note that systematic uncertainties likely influence both results. For example, our model result uses comparatively low luminous efficiencies; adopting those used by other investigators (eg. Borovička et al. 2013a) would tend to reduce our mass and hence energy estimate by as much as a factor of three. Similarly, the period estimate from Hedlin et al. (2018) is likely associated with one of the terminal flares (as they note), which would necessarily represent a smaller total energy than the initial event energy. Given these uncertainties, we consider these independent estimates to be essentially in agreement and collectively they suggest the total mass of Hamburg is of the order tens to at most a few hundred kilograms.

A final energy comparison/estimate may be made using the relationship between peak brightness and total energy found among US Government sensor bolides presented by Gi et al. (2018) (their equation 5). Using our GLM and video lightcurve result that the brightest flare for Hamburg had a peak absolute magnitude of -16.3, we find that the Gi et al (2018) relation predicts an integrated total energy of 7 T TNT equivalent, in excellent agreement with our other estimates.

Considering the lightcurve, ablation modeling, infrasonic energy estimate and energy/magnitude relation together, our preferred total energy range for the Hamburg fireball is 2 – 7 T TNT (8.4-28×$10^9$ J), which corresponds to a preferred mass range of 60 – 225 kg. We consider the upper end of this range the most probable. This implies that the Hamburg meteoroid had an initial diameter in the range 0.3 – 0.5 m.

## Orbital Evolution History and Comparison to other H-Chondrite orbits

In estimating the escape route (ER) of the Hamburg meteorite from the asteroid belt or an ecliptic cometary source we follow the approach described in Granvik and Brown (2018), which utilizes the NEO population model by Granvik et al. (2016). Using the semi-major axis, eccentricity, and inclination as reported in Table 3 (and including the Monte Carlo fit uncertainties, which we emphasize are likely lower limits) we find that Hungaria, Phocaea, and the 2:1J MMR complex have negligible ER probabilities (<<1%) whereas the 3:1J MMR complex (36±8%), Jupiter-family-comet (JFC) region (35±14%), the 5:2J MMR complex (16±8%), and the $\upsilon_6$ secular resonance (12±3%) have significant probabilities. The most striking feature of these probabilities is that there is not a single ER that would clearly stand out. This is partly due to the fact that the orbit of Hamburg and many other H chondrites fall in the densest region of the NEO steady-state orbit distribution which is fed roughly equally by multiple ERs.

The probability for the 5:2J complex is reduced to about 1% and the JFC probability comparably increased (56±8%) if one changes the NEO model by Granvik et al. (2016) to the model by Granvik et al. (2018). Probabilities for the other ERs remain statistically unchanged. The difference between the two NEO models is that the disruption at small perihelion distance ($q$) was modeled differently – Granvik et al. (2016) excluded test particles that reached a critical $q$ when constructing steady-state orbit distributions whereas Granvik et al. (2018) constructed steady-state orbit distributions without considering the disruption and instead used a penalty function (with fitted parameters) to discard the predicted excess of NEOs with small $q$. Hence a small difference in steady-state orbit distributions may explain the difference in ER predictions. This interpretation is also supported by the fact that the probability predictions for JFC and 5:2J complex have the largest error bars when using the Granvik et al. (2016) model and agree on the 2-$\sigma$ level with 56% for JFC and 1% for the 5:2J complex.

The Hamburg meteorite orbit adds evidence to a mid-to-outer belt source region for H-chondrites although an escape through $\upsilon_6$ cannot be excluded (Fig. 15). The two previously known

H4 chondrites (Buzzard Coulee and Grimsby) are likely to originate in the inner asteroid belt or the Hungaria region, albeit from a higher inclination source; Hamburg is the first low inclination H4 orbit. The Hamburg orbit and, consequently, ER probabilities are most similar to the Kosice, Ejby and Mason Gully meteorites, all classified as H5. We note that Hamburg is the fourth H chondrite with a significant probability for an origin in the JFC region, although this may just be a reflection of the overlapping contributions from multiple ERs in this part of the orbital space.

The small orbital inclination of Hamburg would seem to indicate that asteroid (6) Hebe with an inclination of about 15 degrees is unlikely to be the source for H chondrites, particularly as 7 (including Hamburg) of the 14 known H chondrite orbits are under 5 degrees inclination. However, this is not a strong conclusion as a fraction of test particles that start on high-inclination orbits in the asteroid belt may evolve to low-inclination orbits in the near-Earth space (Granvik et al., 2017). Since meteoroids on low-inclination orbits are more likely to impact the Earth compared to those on high-inclination orbits, one would expect the sample of meteoroids and meteorites coming from Hebe to be positively biased towards low-inclinations. Thus, while we cannot conclusively identify the source body of H chondrites, the data at hand suggests that a mid-to-outer belt source region for H chondrites is likely and a parent at low inclination would be favored. Nevertheless, we emphasize that (6) Hebe is located in this region of the asteroid belt and so cannot be ruled out, provided some daughter fragments from Hebe migrate to low inclinations and there is a large enough reservoir of such bodies.

## Conclusions

The Hamburg fireball entered the atmosphere at 15.83 ± 0.05 km/s at a steep angle (only 24 degrees from the vertical) from a radiant to the ESE of Hamburg Lake, MI resulting in a comparatively compact strewnfield oriented almost East-West. Based on darkflight modelling, comparison to meteorite recovery locations and masses together with interpretation of the Doppler radar signature of the falling debris curtain, most recovered fragments originated near the terminal point of the fireball. Some larger fragments may have been separated at the flares to reach darkflight near 20 km altitude, but none of the recovered meteorite masses/locations are consistent with an origin much higher than the first flare. As our entry modelling and the Doppler radar signature suggests a large number of small (<100 g) fragments were produced, the darkflight results predict material in the tens of gram range from the flares would not be primarily found on the lake surfaces, but rather on land between Strawberry Lake and Hamburg Lake where few fragments were recovered. Lack of material from these flares may be due in part to the more difficult search conditions in this region.

The video-derived lightcurve and GLM-derived lightcurve show relative timing and magnitude agreement, though the GLM recorded only the two brightest flares. These flares are centred at heights of 24.1 km and 21.7 km respectively, where the dynamic pressure was 5-7 MPa, suggesting this was the global strength of the bulk of the Hamburg meteoroid. Early minor fragmentation may have occurred near 0.3 MPa, but the evidence for this is not conclusive. That the Hamburg fireball lightcurve derived from video data is consistent supports the claim of Jenniskens et al. (2018) that the GLM sensitivity is near an absolute magnitude of -14.

The Hamburg meteoroid had a pre-atmospheric diameter between 0.3-0.5m, equivalent to an initial mass in the 60 – 225 kg range or a total fireball initial kinetic energy of 2-7 T TNT. Objects the size of Hamburg impact Earth daily (Brown et al. 2002). Based on its pre-atmospheric orbit, the Hamburg H4 chondrite originated in the mid-outer belt. No specific parent body can yet be identified for the H chondrites based on available pre-atmospheric orbits, but the low inclination of Hamburg together with half of all known H-chondrite orbits favors a low inclination parent in the mid to outer belt, though (6) Hebe cannot be strictly ruled out.

**Acknowledgements**

Funding support for this work was provided by the NASA Meteoroid Environments Office through co-operative agreement NNX15AC94A. PGB thanks the Canada Research Chair program, and the Natural Sciences and Engineering Research Council of Canada for funding support. MG acknowledges funding by the Academy of Finland (grants 299543 and 307157). We particularly wish to thank the following videographers for contributing their fireball video records for this work: the Baden Family, Mark Frey and Diana MacPherson, Owen Wolter and Kevin McQuad Jr. at WindsoriteDOTca News, Andrea Mercatante and Michael Kisser at Railstream LLC, John Lalande and Fred Best from the University of Wisconsin-Madison Space Science and Engineering Center, Daniel McGhee, David Barker, Brian S. Bachert, Tom Masterson, and the Kinross Corectional Facility of the Michigan Department of Corrections. Thanks also to Kevin McGrath at the NASA Short-term Prediction Research Transition Center (SPoRT) for providing initial quick-look GLM data and cloud imagery. Additionally, we thank Todd Slisher, B. Barnibo, B. Wolfe, Emily Licata and Andrew Licata for aiding in our search for meteorites. We also thank the government of Hamburg Township for allowing us to search public lands in support of our scientific study. Finally, we to thank Brandon Weller, Ron Matthews, Lisa Matthews, Ashley Moritz, Chris Rodgers, Luke Janes, David Eagen and others that wishes to remain anonymous for sharing the locations and masses of the meteorites that they found. The FM model code was kindly provided by Z. Ceplecha, P. Spurny and J. Borovička and the MILIG code by J. Borovička. Helpful discussions with W.F. Bottke are gratefully acknowledged. We also thank Professor David Sept of the Center for Computational Medicine and Bioinformatics at the University of Michigan for bringing this fireball to our prompt attention. We thank G. Tancredi and P. Spurny for careful reviews of an earlier version of this manuscript. Very helpful discussions with J. Borovička and P. Spurny related to trajectory measurement uncertainty in particular are gratefully acknowledged.

## Tables.

Table 1. Details of videos used in astrometric and photometric measurements of the Hamburg fireball. Station numbering given in the figures is shown in square brackets under the video location where applicable. Videos used for Astrometry [A], Photometry [P] also indicated after location name.

| Video Location[URL] | Latitude, Longitude (N/W) [degs] | Frame Rate (frames per second) | Duration of fireball signal (sec) | Field of View (HxV) [degs] | Sensor resolution (HxV) [pixels] | Range to endpoint (km) |
|---|---|---|---|---|---|---|
| Ypsilanti, MI[1] [P] [27] | 42.27, 83.60 | 15 | 2 | - | 1920x1080 | 44 |
| Defiance, OH[2][A,P] [95] | 41.24, 84.36 | 15 | 4.1 | 70x35 | 1200x600 | 155 |
| Oberlin, OH (NASA All-Sky) [P] [16] | 41.29, 82.22 | 30 | 4.4 | 360x90 | 640x480 | 190 |
| Brant, ON[3] [A] [99] | 43.21, 80.23 | 1 | 4 | 84x51 | 2688x1520 | 307 |
| Chicago, IL[4] [A] [2] | 41.86, 87.64 | 30 | 4.1 | 60x24 | 1500x600 | 320 |
| Madison, WI[5][A,P] [1] | 43.07, 89.41 | 7.7 | 3.4 | 76x47 | 1280x960 | 453 |

[1]https://www.youtube.com/watch?reload=9&v=mu0BpkFSPJU

[2]https://twitter.com/BadenElizabeth/status/953745866518560770

[3]https://www.youtube.com/watch?v=jHIu_Kz3srU

[4]https://www.youtube.com/watch?v=2OEw7YkyXKQ

[5]https://www.youtube.com/watch?v=8Pf5739vHoU

Table 2. The atmospheric trajectory for the January 17, 2018 Hamburg fireball based on four calibrated camera measurements. Geographic coordinates are referenced to the WGS84 geoid and all local quantities are in an Earth Centred Fixed (ECF) frame. Uncertainties represent the best overall estimate of the error in speed and radiant. Also shown are error bounds (in square brackets) that represent the uncertainty in mathematical fit (derived from the Monte Carlo uncertainty as described in the text) as a lower error estimate and the standard deviation of the interstation speed as an upper error estimate.

| | Beginning | End |
|---|---|---|
| Height (km) | 83.02 ± 0.01 | 19.73 ± 0.01 |
| Latitude (N) | 42.320° ± 0.0001° | 42.451° ± 0.0001° |
| Longitude (W) | 83.567° ± 0.0005° | 83.857° ± 0.0002° |
| Slope | 66.14° ± 0.29° [0.02, 0.29] | |
| Azimuth of radiant | 121.56° ± 1.2° [0.13, 1.2] | |
| Velocity (km/s) | 15.83 ± 0.05 [0.01, 0.14] | < 6 |
| Trail Length/Duration | 68.7 km />4.2 s | |
| Time (UT) | 01h08m29s | 01h08m34s |

Table 3. Heliocentric orbit for the fireball producing the Hamburg meteorite. All angular coordinates are referenced to J2000.0, except the apparent radiant which is epoch of date and in an Earth Centred Fixed (ECF) frame. $V_\infty$ refers to the speed of the fireball relative to the Earth's surface prior to significant atmospheric deceleration, which for Hamburg occurs at a height of 50 km. Uncertainties represent the best overall estimate of the error in speed and radiant. Also shown are error bounds (in square brackets) that represent the uncertainty in mathematical fit (derived from the Monte Carlo uncertainty as described in the text) as a lower error estimate and the standard deviation of the interstation speed as an upper error estimate.

| | |
|---|---|
| $\alpha_r$ | 72.83 ± 0.34° [0.04, 0.34] |
| $\delta_r$ | 27.37 ± 0.30° [0.03, 0.30] |
| $V_\infty$ | 15.83 ± 0.05 km/s [0.01, 0.14] |
| $V_G$ | 11.1 ± 0.07 km/s [0.01, 0.2] |
| $\alpha_G$ | 74.29 ± 0.39° [0.05, 0.4] |
| $\delta_G$ | 24.71 ± 0.36° [0.03, 0.36] |
| a | 2.73 ± 0.05 [0.01, 0.11] A.U. |
| e | 0.661 ± 0.006 [0.001, 0.014] |
| i | 0.604 ± 0.11° [0.01, 0.11] |
| ω | 211.65 ± 0.3° [0.03, 0.3] |
| Ω | 296.421± 0.03° [0.003, 0.04] |
| q | 0.926 ± 0.001 [0.002, 0.001] A.U. |
| Q | 4.5 ± 0.1 A.U. [0.005, 0.2] |
| $T_j$ | 2.99 ± 0.003 [0.002, 0.01] |

Table 4. Documented Hamburg meteorite recoveries. Locations and mass were directly verified for all specimens; AMS indicates information published by the American Meteor Society at https://www.amsmeteors.org/members/imo_view/meteorites/2018/168 (accessed August 20, 2018).

| Date | Finder | Location | Mass | Notes |
|---|---|---|---|---|
| 18-Jan-18 | T. Slisher | 42.4485996, -83.8358445 | 12g | Found on lake |
| 18-Jan-18 | T.Licata | 42.4488509, -83.8385039 | 15.83g | Found on lake |
| 18-Jan-18 | B Wolfe | 42.4490422, -83.8483335 | 26g | Found on lake |
| 20-Jan-18 | A. Licata | 42.4502360, -83.8589170 | 0.9g | Multiple Specimens |
| 20-Jan-18 | B. Barnibo | 42.4362000, -83.7983910 | 3g | Found on lake |
| 20-Jan-18 | E. Licata | 42.4351030, -83.7943220 | 0.301g | Found on lake |
| 27-Jan-18 | T.Licata | 42.4508390, -83.8226400 | 10.43g | Found in wooded area |
| 18-Jan-18 | D.Landry | 42.4541500, -83.8641120 | 20g | AMS |
| 19-Jan-18 | G. Barger | 42.4533040, -83.8608600 | 11g | AMS |
| 19-Jan-18 | L. Janes | 42.4512560, -83.8603470 | 20g | AMS |
| 26-Jan-18 | R. Matthews | 42.4507690, -83.8589820 | 2g | AMS |
| 26-Jan-18 | L. Matthews | 42.4506450, -83.8588070 | 1g | AMS |
| 18-Jan-18 | A. Larry | 42.4484540, -83.8590680 | 17.5g | AMS |
| 18-Jan-18 | Resident | 42.4500280, -83.8542220 | ~60g | Witnessed by Brandon Weller, found on land |
| 18-Jan-18 | B. Weller | 42.4520280, -83.8504440 | 59.4g | AMS |
| 18-Jan-18 | R. Ward | 42.4511390, -83.8476390 | 102.6g | Largest found fragment - Witnessed by Brandon Weller |
| 18-Jan-18 | L. Atkins | 42.4488730, -83.8386330 | 37g | AMS |
| 20-Jan-18 | L. DeLanoy | 42.4471920, -83.8276960 | 6.5g | AMS |
| 19-Jan-18 | T.V. | 42.4475530, -83.8359690 | 13.8g | |
| 19-Jan-18 | T.V. | 42.4475140, -83.836789 | 12.6g | |
| 19-Jan-18 | T.V. | 42.4470070, -83.8384670 | 11.5g | |
| 22-Jan-18 | D. Grischke | 42.4535890, -83.8560520 | 55.92g | AMS |
| 27-Jan-18 | T. Licata | 42.4478106, -83.8135184 | .2g | Found on Air Field |
| 28-Jan-18 | T. Licata | 42.4511370, -83.8532140 | 0.008g | Found in Baseball Field |
| 20-Jan-18 | A. Moritz | 42.4573020, -83.8473590 | 50g | |
| 18-Jan-18 | Unidentified | 42.4512560, -83.8560530 | 20.6g | |

## Figure Captions

Figure 1. Regional map overview showing cloud cover as measured by the GOES 16 Advanced Baseline Imager (Level 1b data) at 01:02 UT on Jan 17, 2018 in the area where the fireball (red line) was visible. These images are in the 10.35 μm band and are showing longwave IR radiance (color bar) in units of mW/($m^2$ sr $cm^{-1}$) . Credit: NASA SPoRT/Kevin M. McGrath.

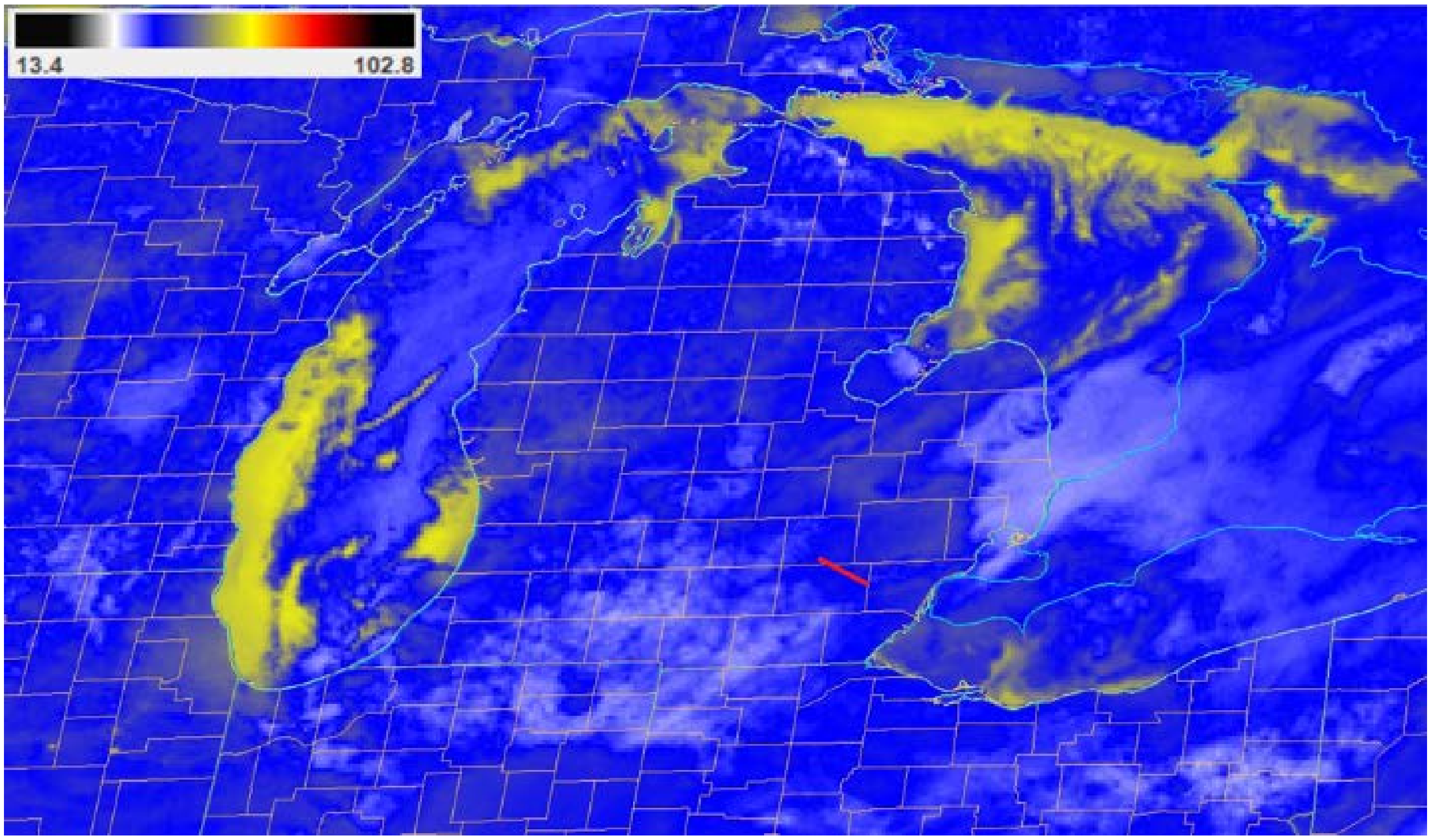

Figure 2. The Hamburg fireball as seen from Oberlin, OH [16] (top left), from Defiance, OH [95] (top right), Chicago, IL [97] (lower left) and from Madison, WI [1] (bottom right). The arrow points to the fireball in each frame with the inset showing a zoomed in region near the tip of each arrow for clarity. Credits: NASA MEO, the Baden Family, Railstream, LLC/Andrea Mercatante and Michael Kisser, and UW-Madison SSEC/AOS, respectively.

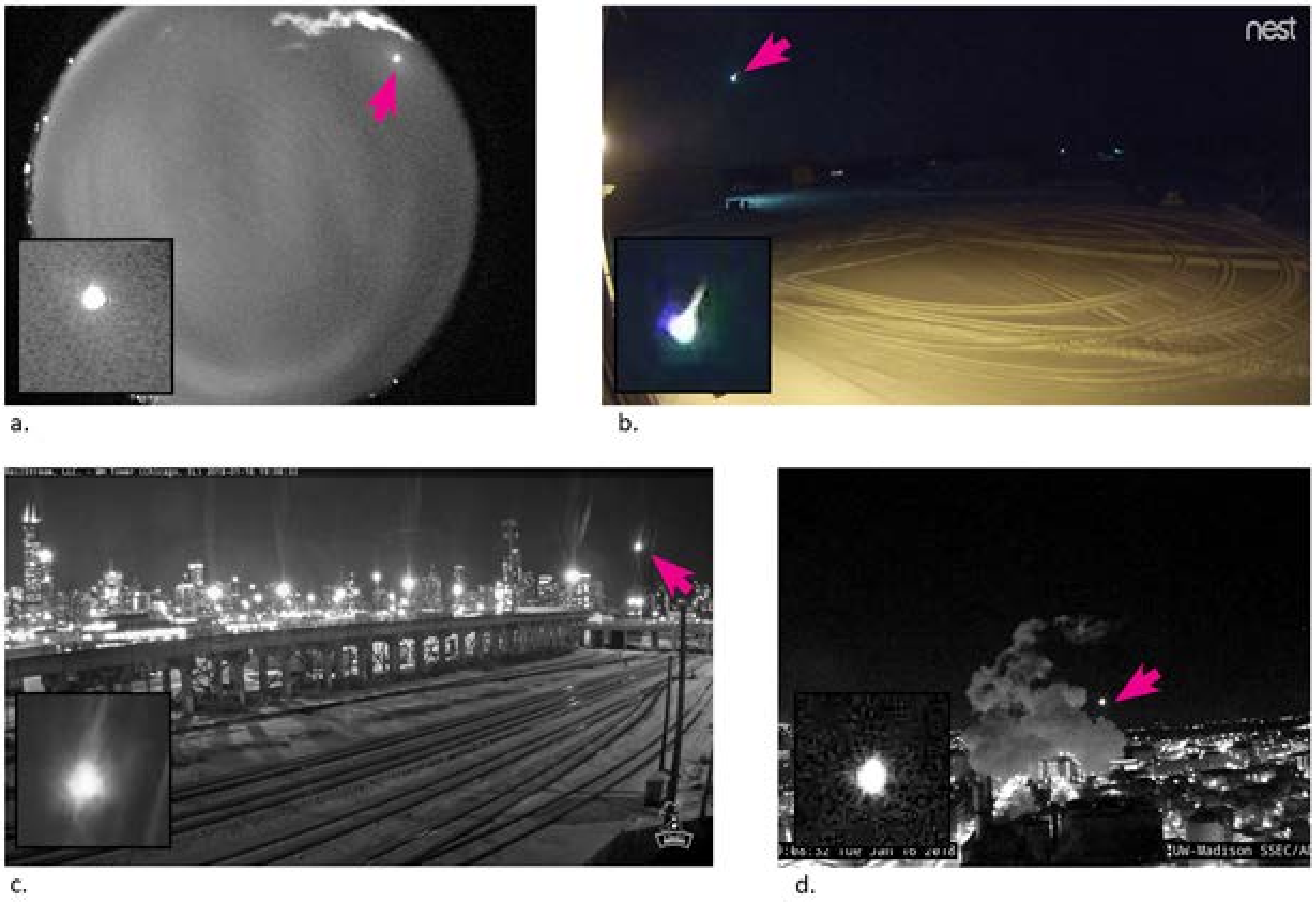

Figure 3. Location of cameras used for astrometric measurements and ground track (red line) of the Hamburg fireball. Station numbering corresponds to names given in Table 1.

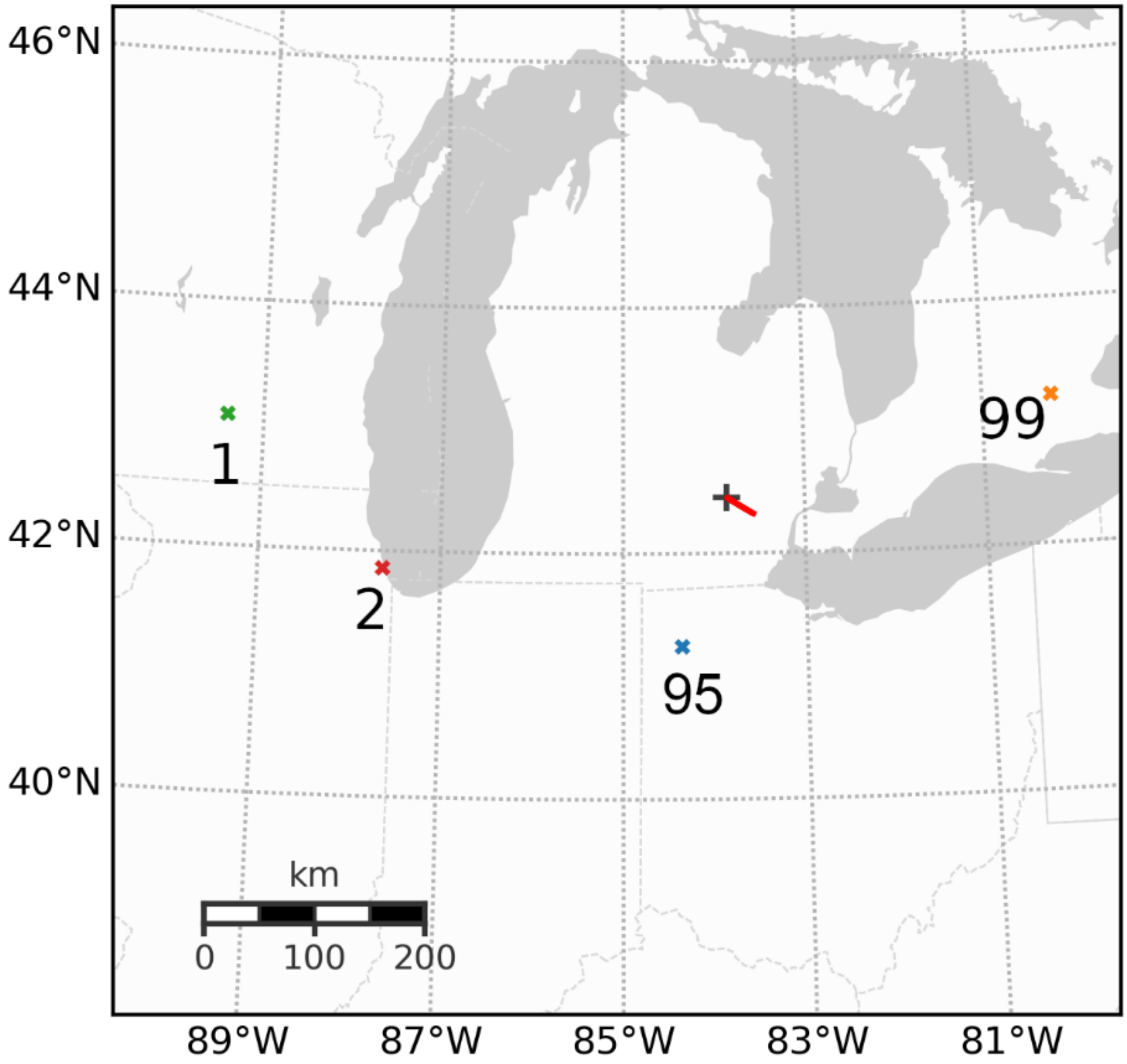

Figure 4. Lateral residuals from all stations relative to the final straight-line solution for the Hamburg fireball. Station numbering corresponds to values in Table 1 with 95 – Defiance, OH, 99 – Brant, ON, Canada, 1 – Madison, WI, and 2 – Chicago, IL.

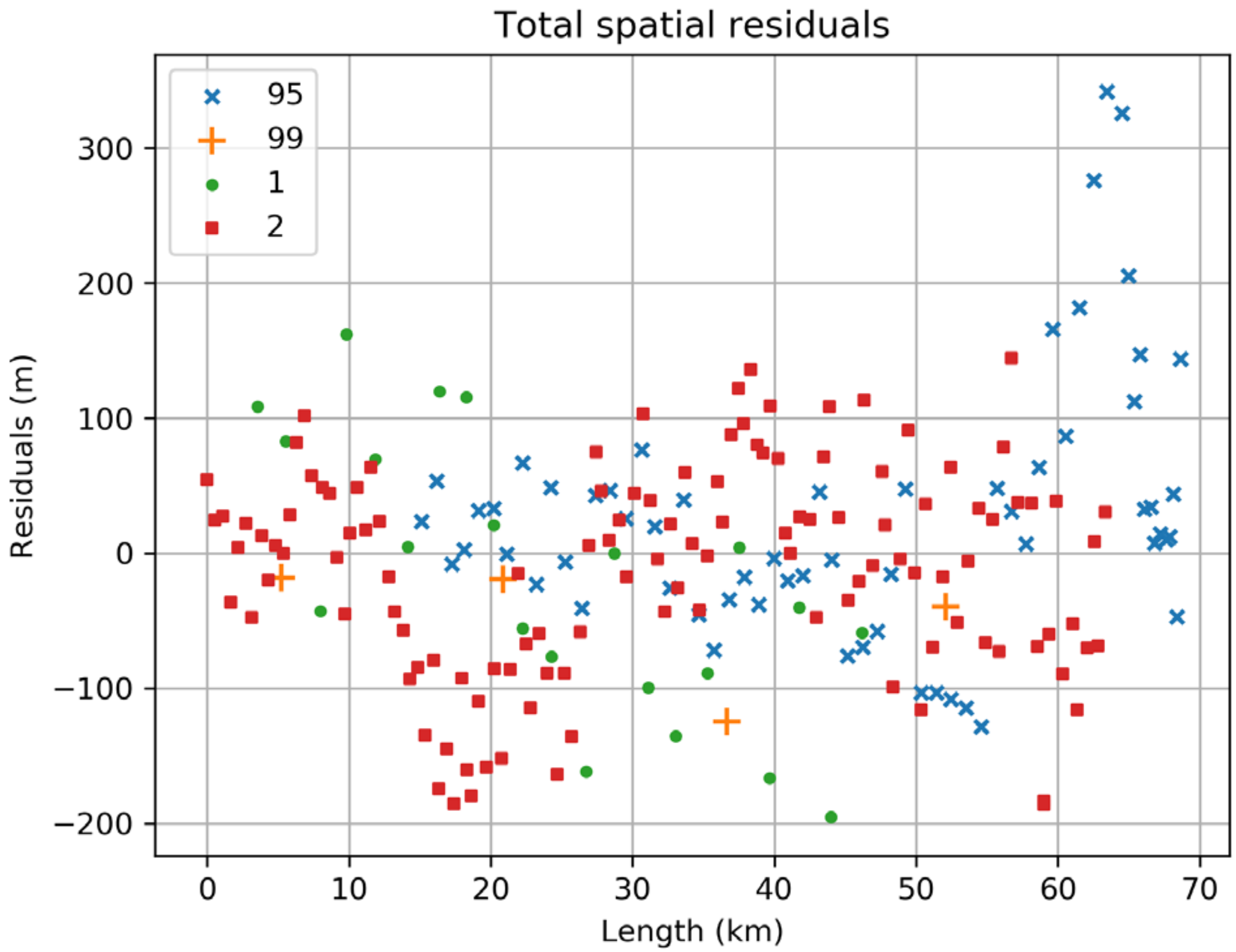

Figure 5. Lightcurve of the Hamburg fireball based on video records. Time t=0 corresponds to Jan 17, 01:08:29.49 UT.

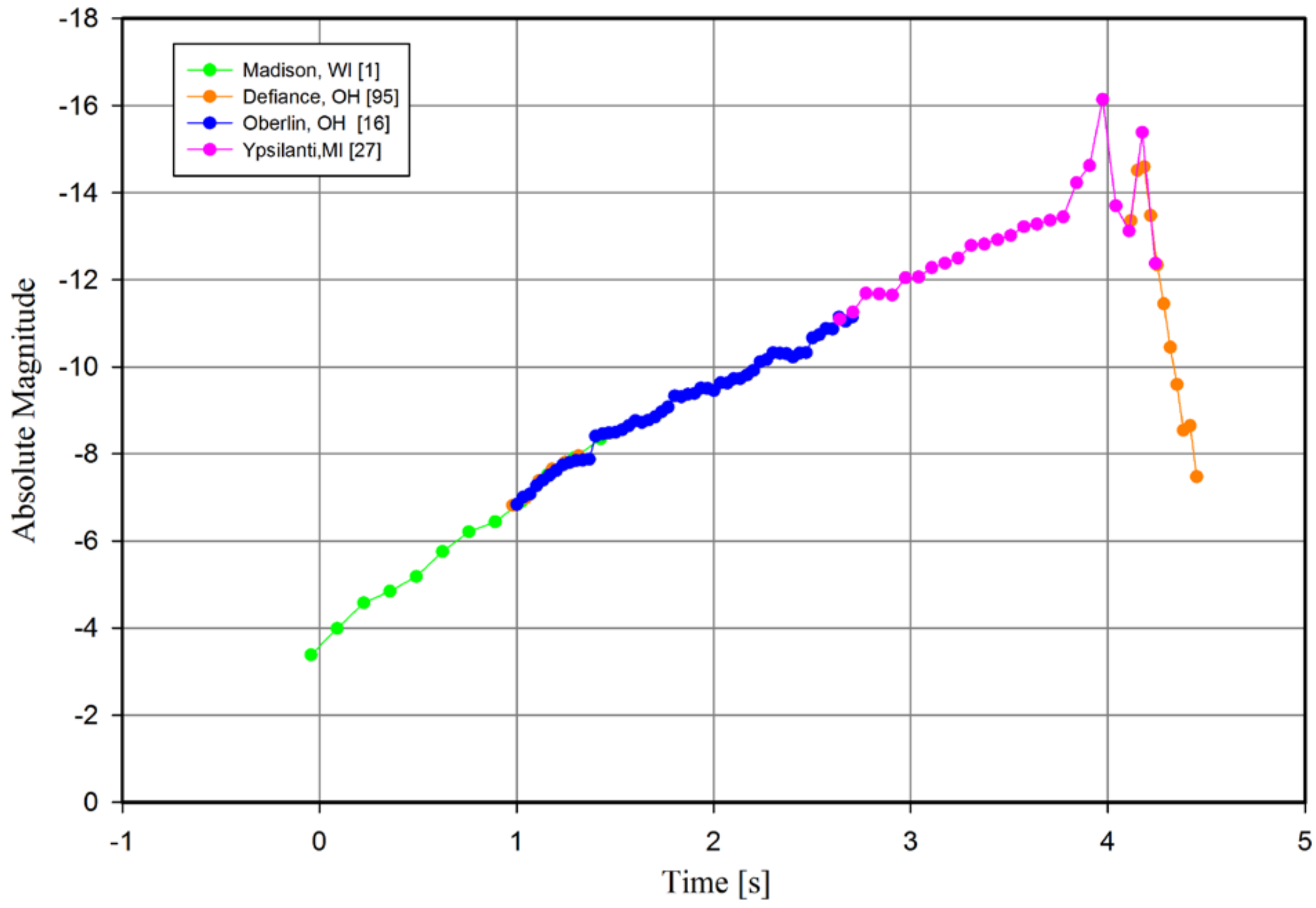

Figure 6. Screen capture of two frames of Ypsilanti, MI video showing regions used for relative brightness (square regions - bottom) changes and calibration lightsources (circled - top). Image credit Daniel McGhee.

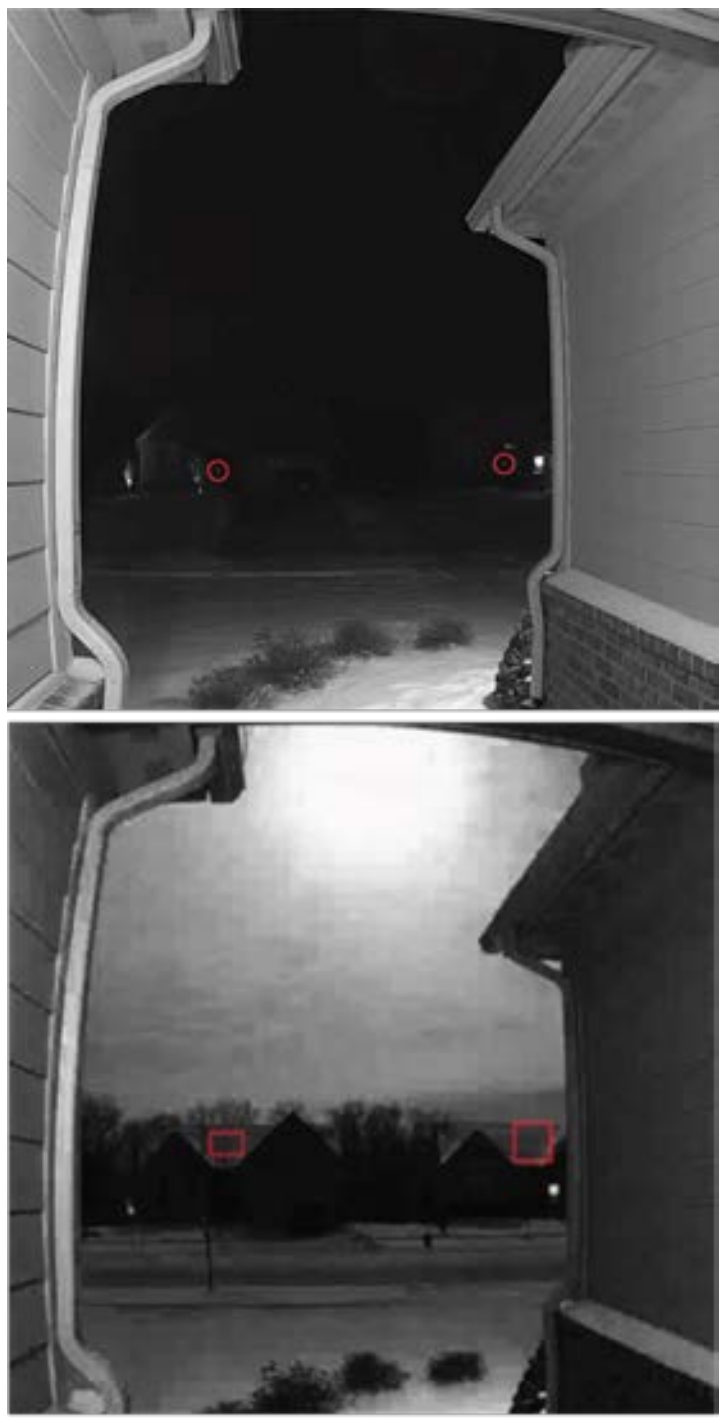

Figure 7. GLM group energy waveform (black dots) in units of $10^{-15}$ J and equivalent spectral energy density (red dots) in units of $\mu Jm^{-2}ster^{-1}nm^{-1}$. Note the drop in the SED from 4.14s to 4.18s is due to the optical groups having two events each, leading to a doubling in the equivalent solid angle of the fireball and an equivalent reduction in the SED.

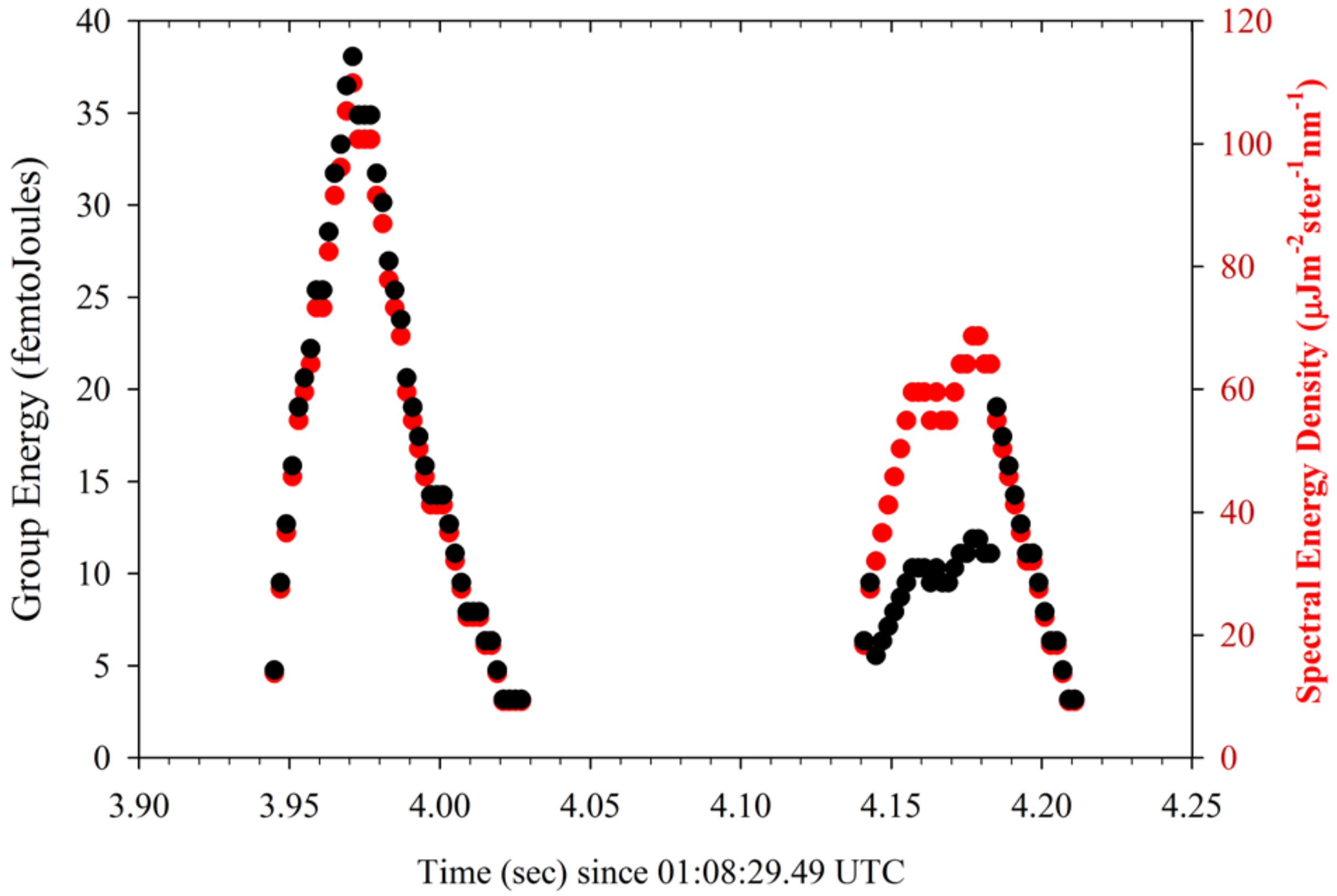

Figure 8. GLM equivalent spectral energy density (red dots) scaled to absolute astronomical magnitude assuming a noise floor at 15 $Wm^{-2}ster^{-1}nm^{-1}$ is equivalent to $M_v$=-14 compared to the video-derived lightcurve (black dots and line).

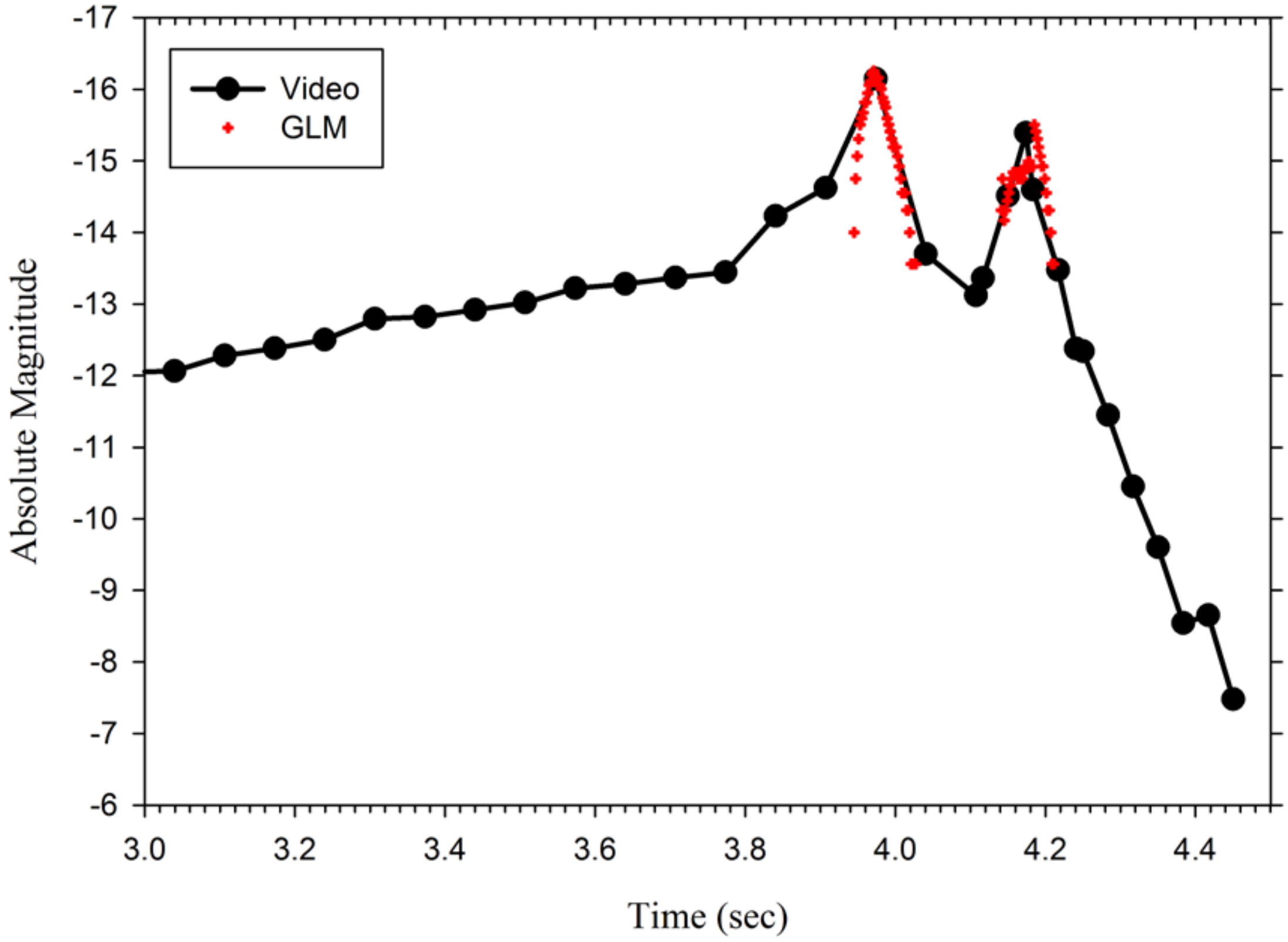

Figure 9. Recovery locations of meteorites (with measured masses) relative to fireball trajectory (red). Also shown are the location of the first flare at 24.1 km altitude (green square), the second flare at 21.7 km altitude (yellow square) and the end point at 19.7 km (purple square).

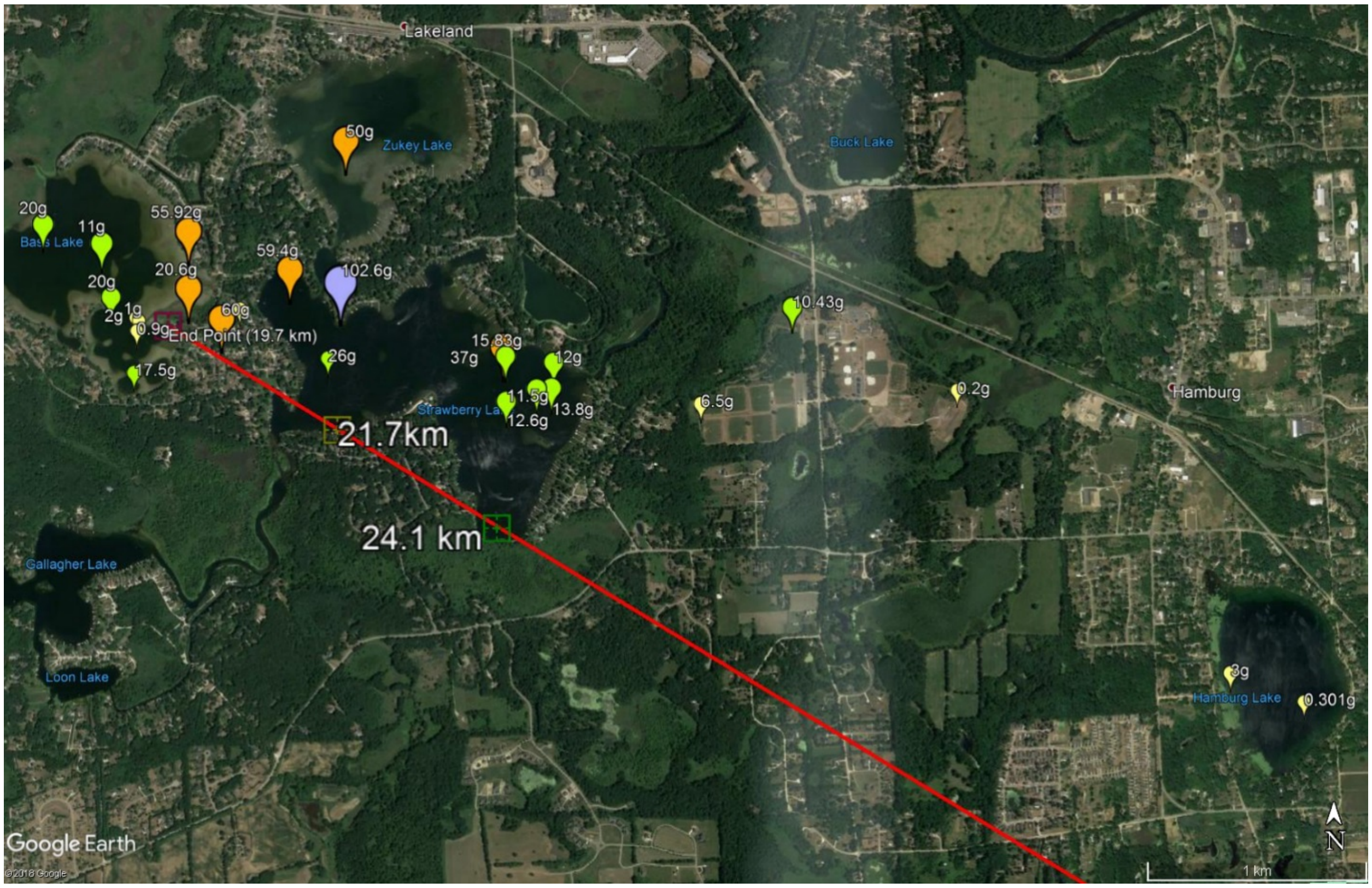

Figure 10. Upper winds measured by radiosonde released at White Lake, MI (https://www.ncdc.noaa.gov/data-access/weather-balloon-data). The top left plot shows the wind field at 0 UT on Jan 17 while the bottom is the wind field at 12 UT, Jan 17.

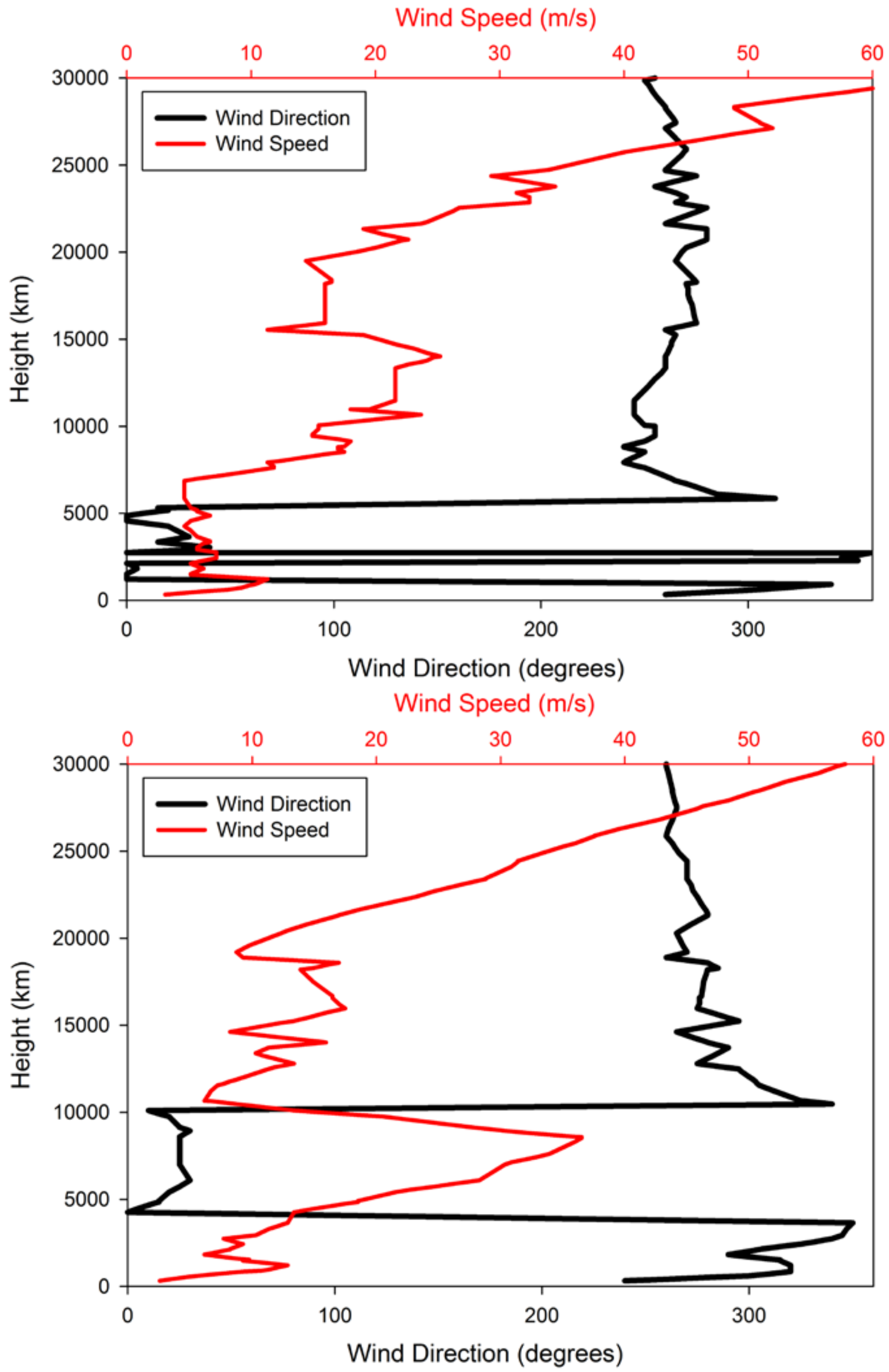

Figure 11. Darkflight model predicted meteorite fall locations. Predicted fall points at the ground for masses of 1 kg, 100g, 50g, 20g, 10g and 1 g are shown released from the end point and at the height of the first flare, with the symbols color coded to match the endpoint or first flare.

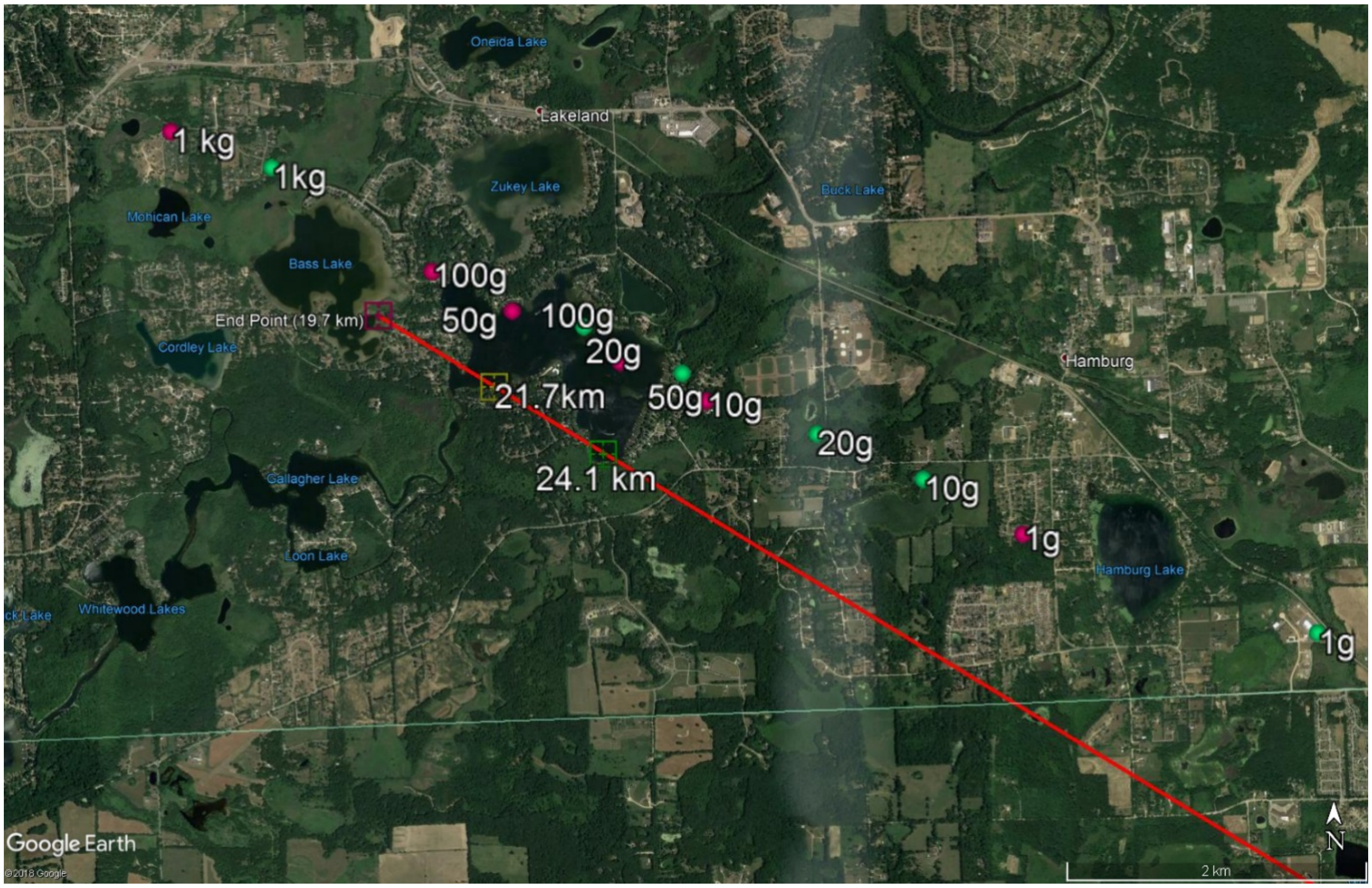

Figure 12. Dispersion in darkflight fall locations for Monte Carlo spreads of 50 m/s for fragments released from the fireball end point for masses of 1kg, 100g, 10g and 1g respectively (from west to east).

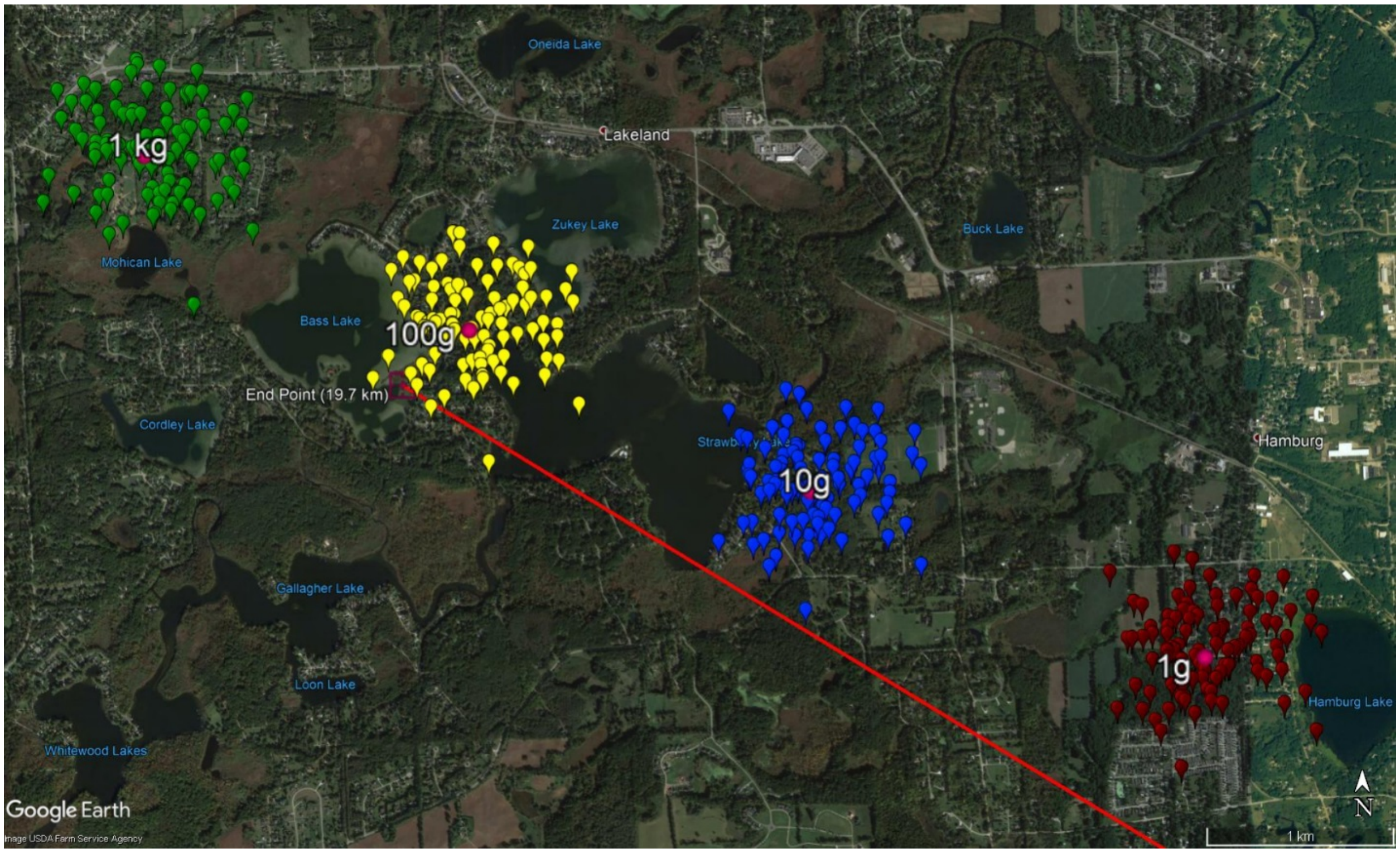

Figure 13. Doppler radar reflectivity returns from t=288sec to t=730 sec after the Hamburg fireball. Note the progression of returns to the East in time reflecting the mass sorting where smaller masses take longer to fall and are blown progressively further East by the prevailing winds. The sweep height in km is shown for the center of each plot as is the radar signal return strength colorscale (lower right). Shown are the individual best fit darkflight model matches in location and time for various masses released at different heights. Masses released at the endpoint are shaded purple, those from the first flare are green. The mass and modelled time the fragment fell through the radar beam in seconds either before the beam sweep (negative) or after (positive) are also shown.

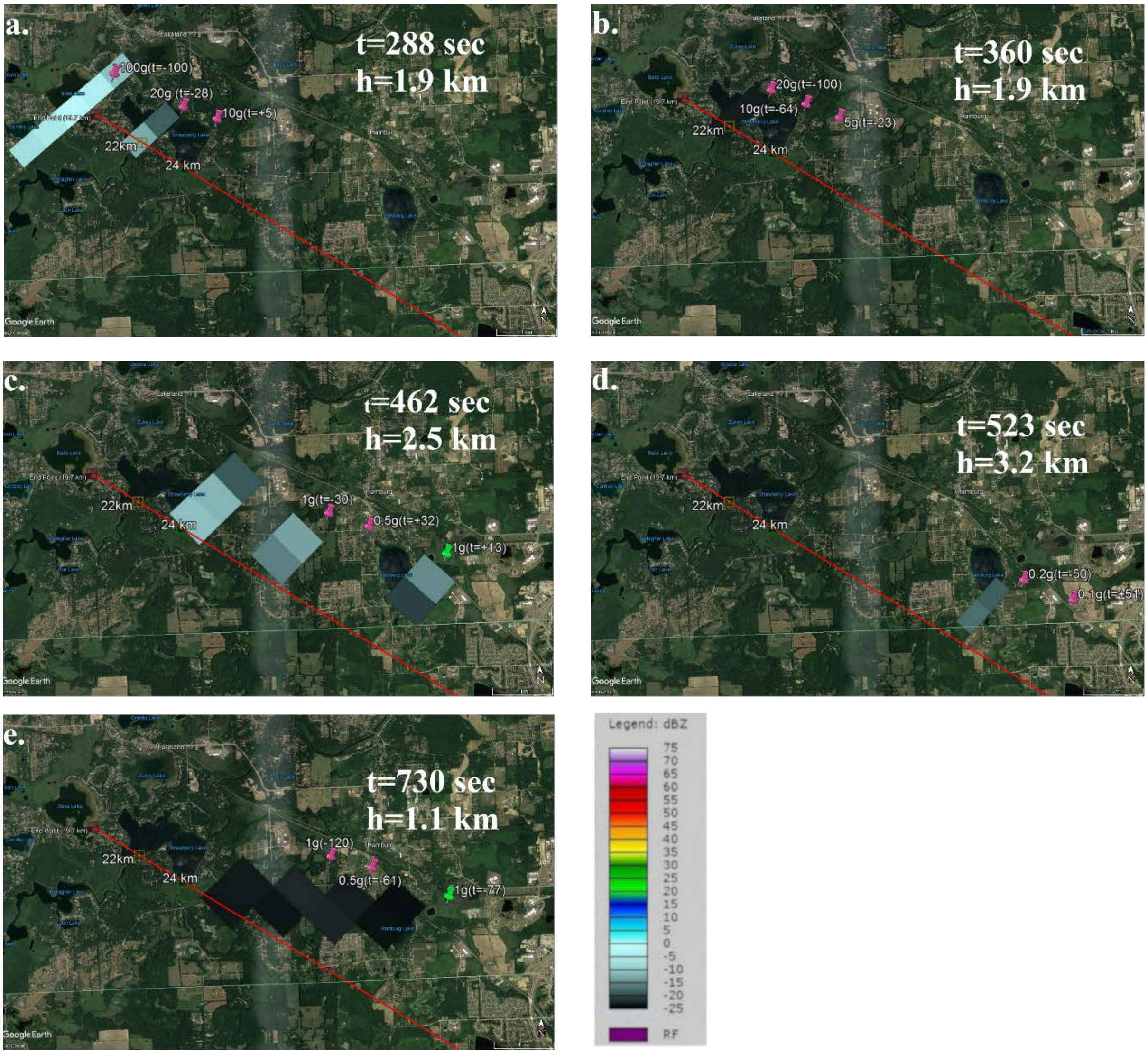

Figure 14. FM model fit (red line) to observed Hamburg fireball lightcurve (black dots) as a function of height.

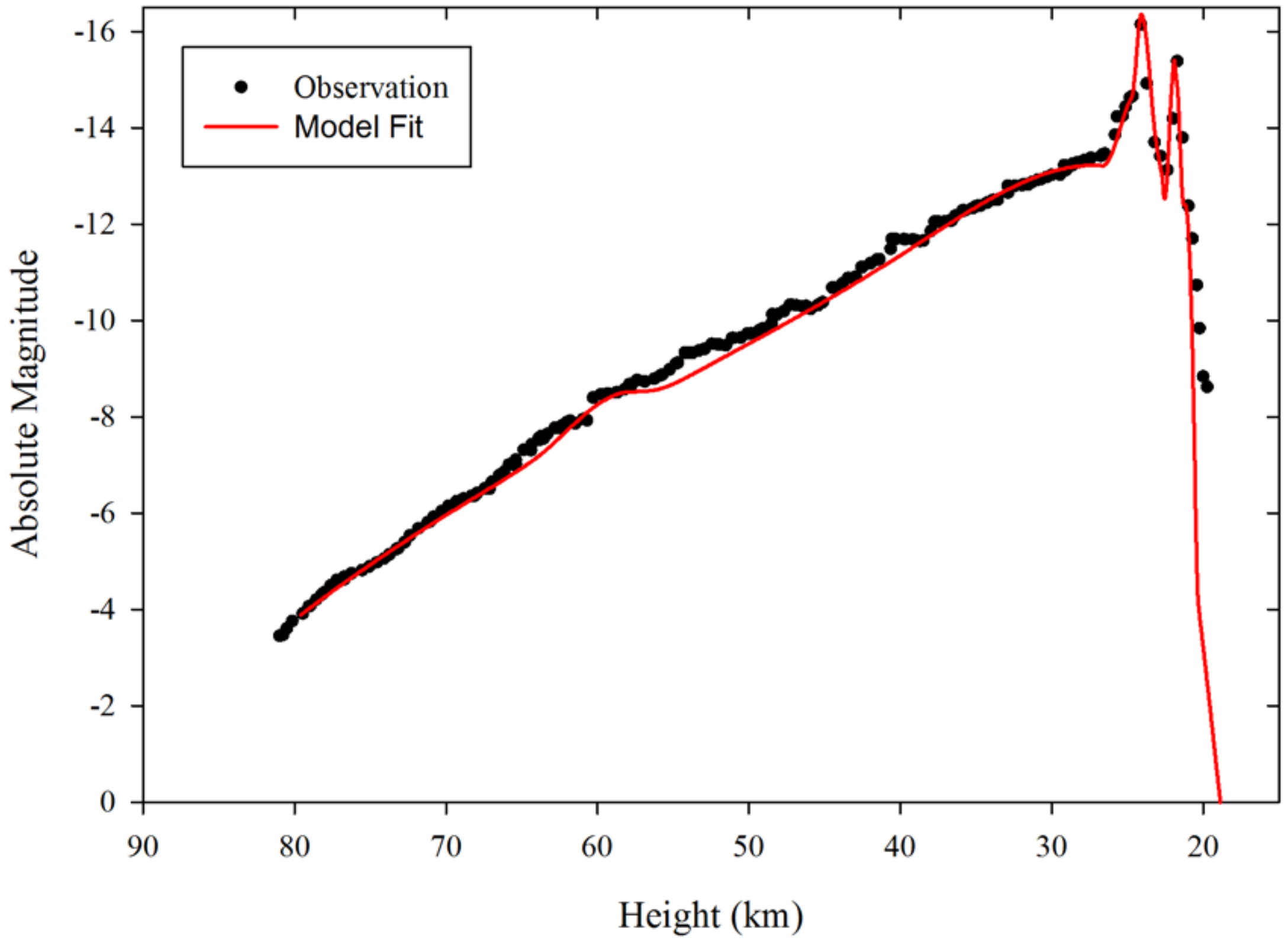

Figure 15. Escape-route probabilities for known H chondrites including the Hamburg meteorite. The symbol size is proportional to the probability that the measured meteorite orbit at Earth originated from a given escape route in the main belt (see Granvik et al., 2016 and Granvik et al 2018 for details of the model).

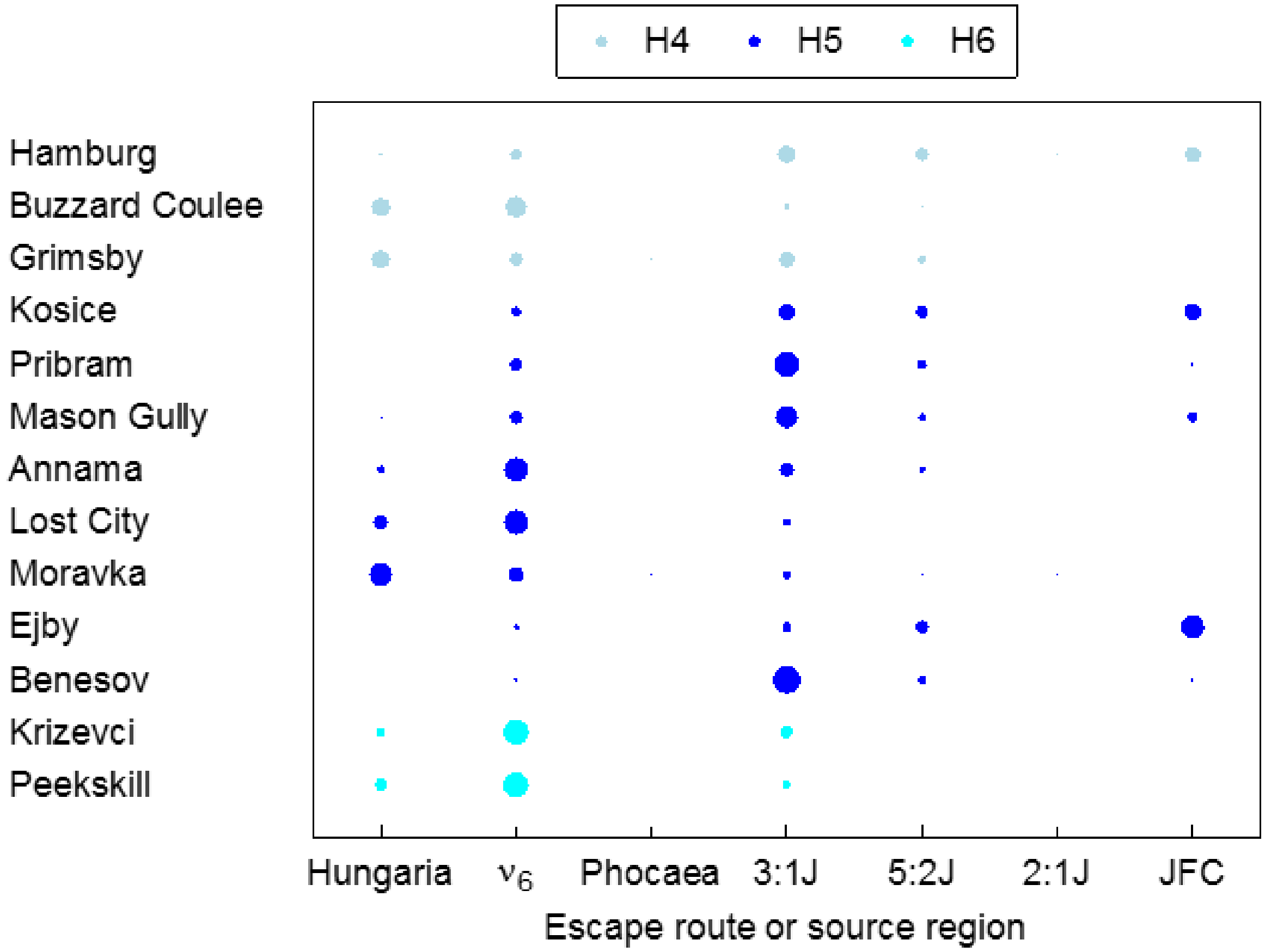

# Appendix A: Calibration details and Trajectory Solution for the Hamburg fireball

In what follows we present details of the stellar calibrations, plate fits and comments regarding observing circumstances for the four stations used in the final trajectory solution for the Hamburg fireball together with information about the trajectory solution itself. We refer to these by the station numbers given in Table 1 in the main text. The four calibrations used for the trajectory solution and for orbit determination were Defiance, OH [95], Brant, Ontario [99], Chicago, IL [2] and Madison, WI [1].

Note that a right handed coordinate system is used for local azimuth (phi) and zenith distance (Theta), where phi=0 (East), 90 (North), 180 (West) and 270 (South).

## Station 1 – Madison, MI

The imagery from this camera was taken from video recorded by the East facing camera on the top of the Space Science and Engineering Center of the University of Wisconsin – Madison. It is located 62m above the surface, providing an excellent view of the horizon. The camera is an IQinvision Sentinel 865 (model #IQ865NE-v7) and standard images from the system are available at http://metobs.ssec.wisc.edu/pub/cache/aoss/cameras/east/latest_orig.jpg

A calibration plate was made using stacked video (100 frame stacks) using 110 individual star picks (some of the same star but at different times) from the camera on nights between Jan 5-21, 2018 where bright stars were visible in the image. Distant horizon objects were checked on calibration nights to compare with the pixel location from the night of the fireball. In two instances a small (1-2 pixel) shift was noticed and the image adjusted accordingly. Because of the large zenith distance of the fireball and the low altitude of some of the reference stars refraction was taken into account for this station. The stellar residuals as a function of zenith angle are shown in Figure A1 for a 3rd order polynomial fit in x,y. The average residuals are 0.03 degrees. Steam exhaust obscured some of the final portion of the fireball flight.

Figure A2 shows the resulting plate with an overlay of 1x1 degree altitude/azimuth gridding. Also shown is the final trajectory plane for the fireball (blue line), its begin and end points and the true horizon (purple line). Individual stars used from the original fireball imagery for plate calibration are also shown, with residual arrows to show apparent residual direction relative to the plate coordinates. Figure A3 shows the location of the fireball on the video relative to the available star picks used to produce the final plate.

Although the residuals are quite good in this case, we found that using a radial plate fit (Borovička, 2014) produced better agreement in the offset of the final trajectory spatial residuals compared to other stations, producing a change of order a few tenths of a degree at most compared to the 3rd order polynomial fit. We have opted to use the radial fit plate solution for this station alone. These pick points in altitude and azimuth are reflected in the values shown in Table A1.

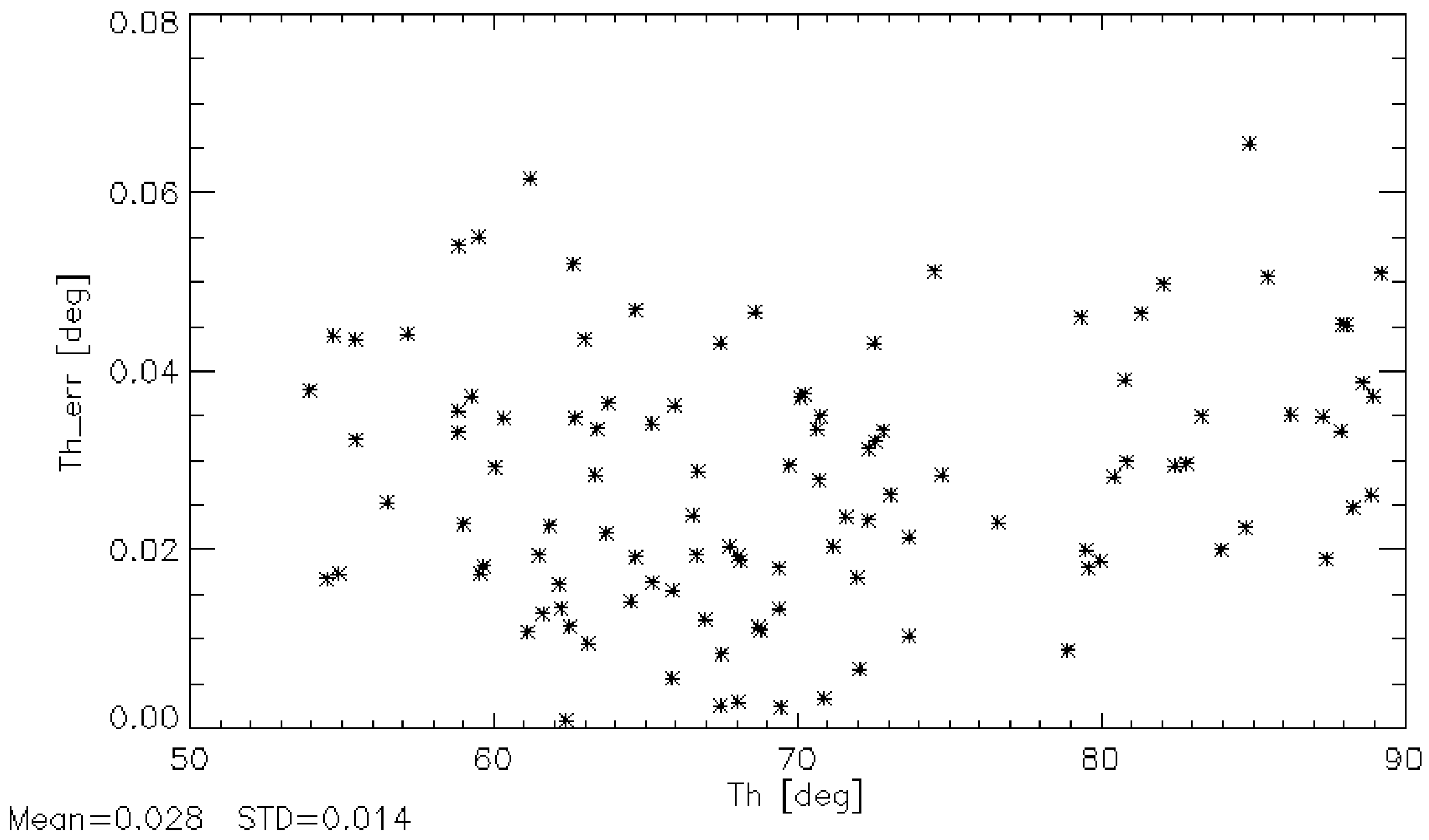

Figure A1. Fit residuals between plate and individual stars used for calibration for station 1 as a function of zenith distance for a 3rd order polynomial plate fit in x,y.

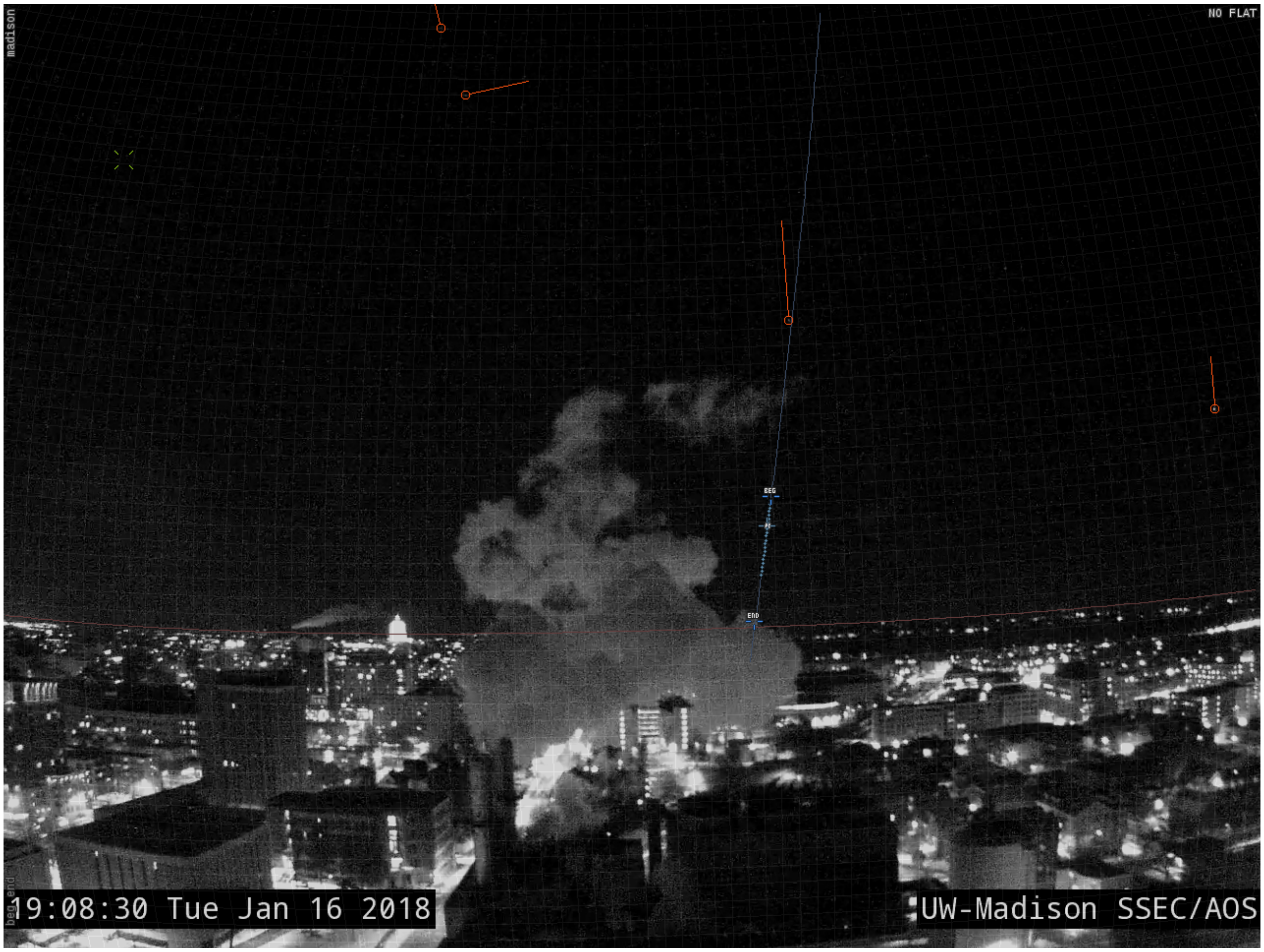

Figure A2. Plate fit for Madison, WI [1] showing altitude/azimuth gird lines in one degree increments. Also shown is the final trajectory plane for the fireball (blue line), its begin and end points and the true horizon (orange line).

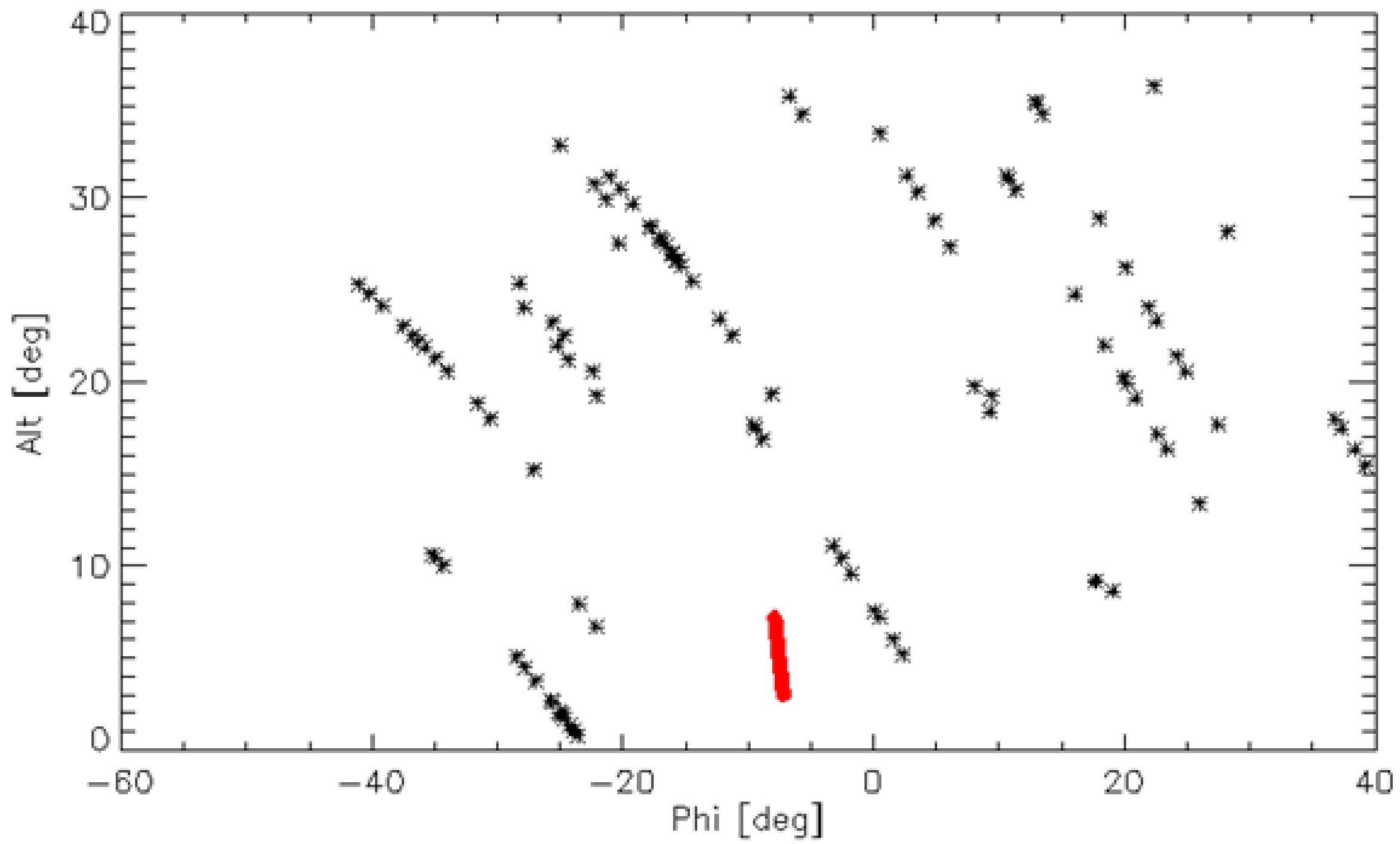

Figure A3. Location of star picks (black dots) relative to the apparent fireball picks (red dots) for Madison, WI.

Station 2 – Chicago, IL

The video from this station was recorded in the city of Chicago as part of the Railstream train monitoring network (https://railstream.net/). Calibration plates were made using streamed video made available from the same camera for times between Feb 1-March 15, 2018 when skies were clear. The highly lit scene made both calibration challenging and fireball picks more complicated than at other sites. A total of 35 individual stellar positions were used in the calibration. Figure A4 shows the pick residuals as a function of stellar zenith distance. The average residuals are again near 0.03 degrees. The earliest frames of the fireball did not permit reliable picks due to compression of the video stream which is most serious for fainter signals.

Figure A5 shows the resulting plate and azimuth/altitude grid overlay while Figure A6 shows the distribution of calibration stars relative to the apparent fireball path in the sky. The bright background lights made getting well placed stellar calibration points difficult.

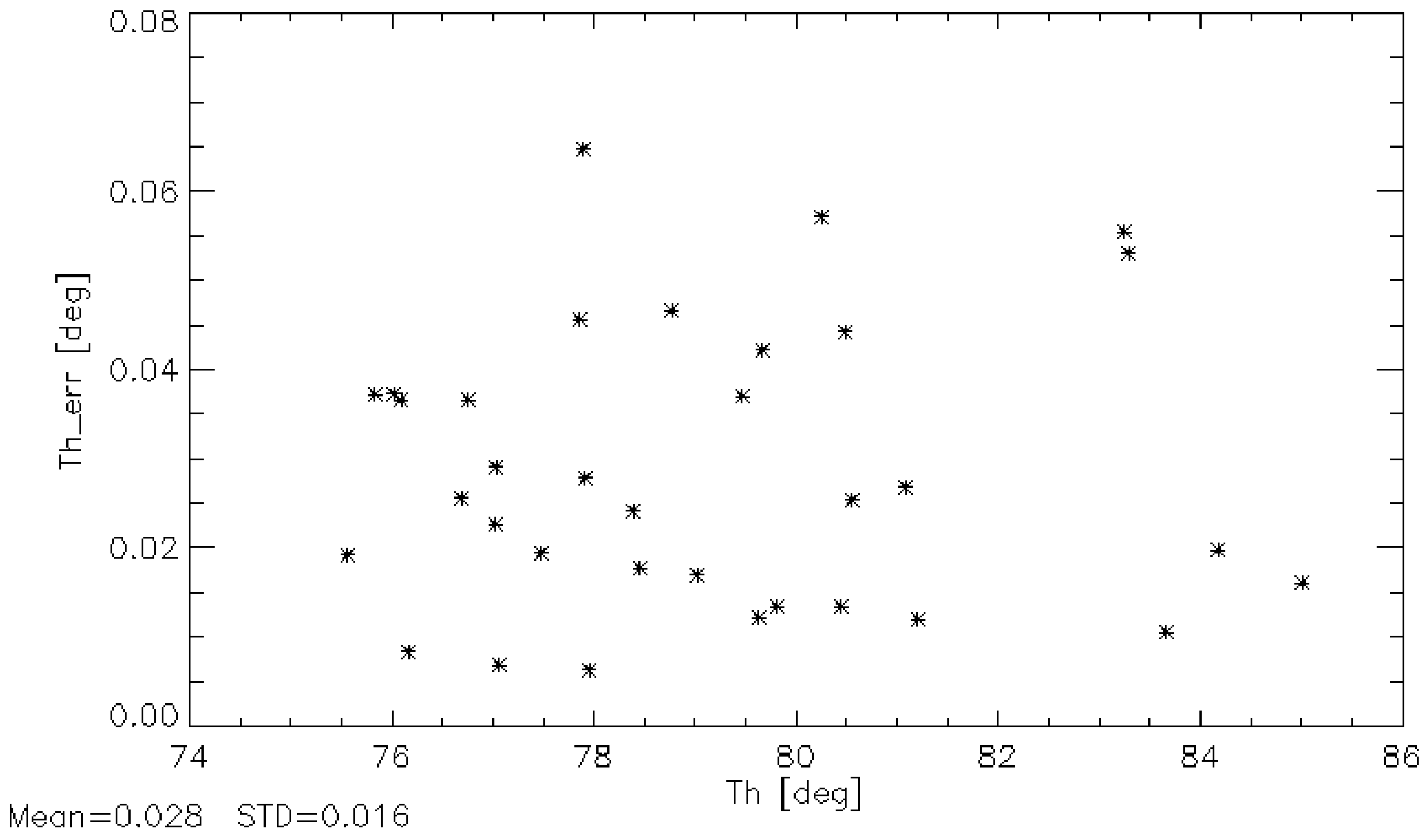

Figure A4. Fit residuals between the plate and individual stars used for calibration for station 2 as a function of zenith distance.

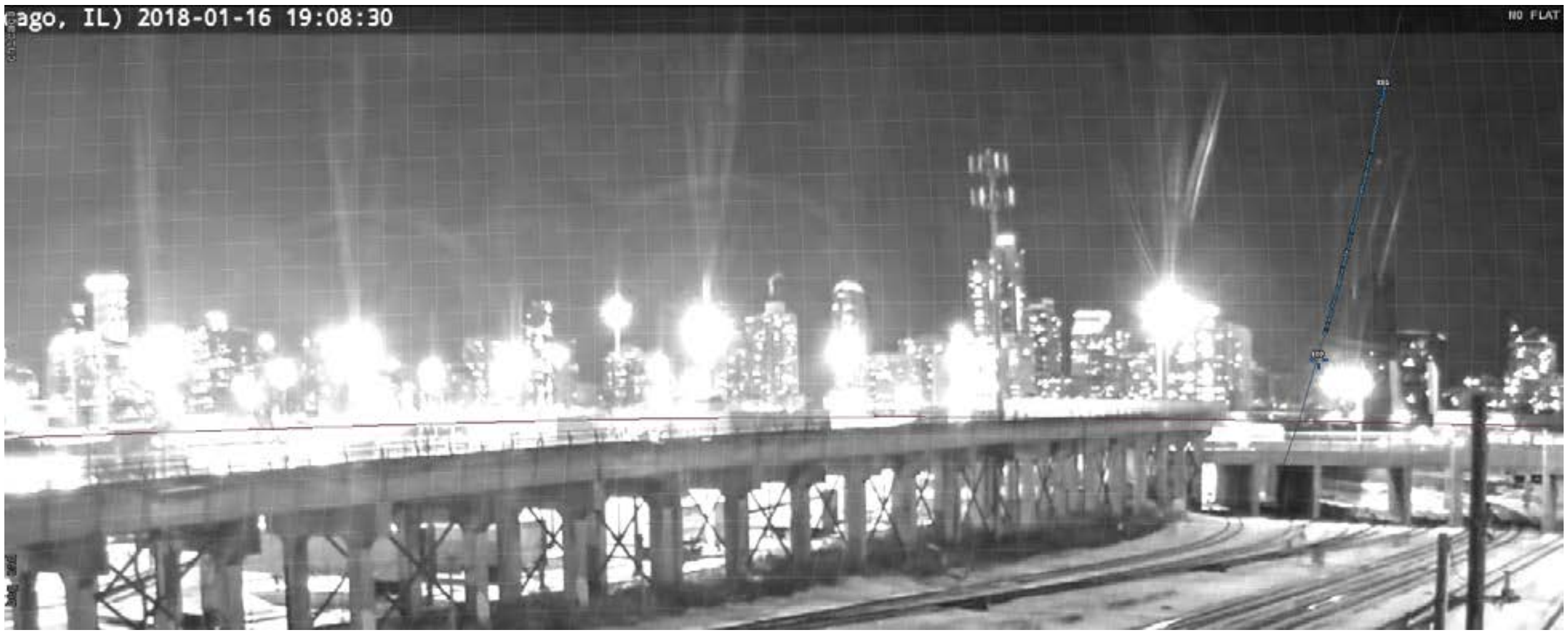

Figure A5. Plate fit for Chicago, IL [2] showing altitude/azimuth gird lines in one degree increments. Also shown is the final trajectory plane for the fireball (blue line), its begin and end points and the true horizon (orange line).

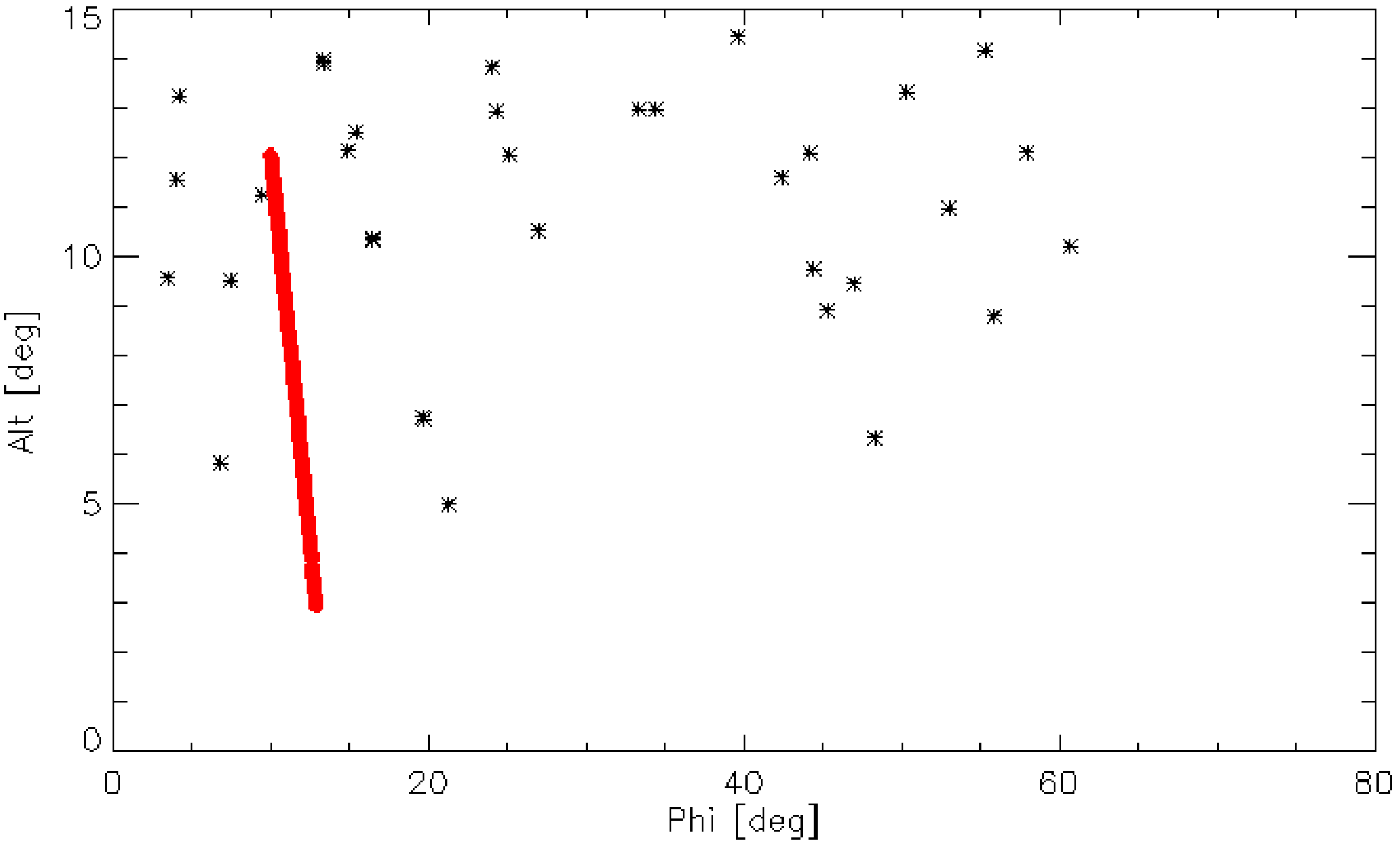

Figure A6. Local coordinates of star picks (black dots) relative to the apparent fireball picks (red dots) for Chicago, IL.

Station 95 – Defiance, OH

This security video was recorded from a private residence in Defiance, OH [95] using a NEST IP camera. A total of 69 stellar calibration positions were secured from stacked video between March 12-31, 2018. Comparing these calibration videos to the fireball video, shifts of up to 3 pixels were apparent on some images, found by comparing common objects on the horizon. These shifts were corrected in the final plate fit. Figure A7 shows the stellar residuals as a function of theta (zenith distance).

From this station the end of the fireball is very near the lower left corner of the field (Figures A8 and A9). It was not possible to find useable stars in any of the available calibration images which overlapped the last few degrees of the fireball path. Distortion in the lower left of the field suggests that systematic offsets may be present in the fireball astrometry near the end of the trail, but with no calibration stars proximal to the fireball endpoint it is difficult to estimate the magnitude of any systematics. The overall stellar fits averaged 0.04 degrees, larger than at other sites. There was also a larger standard deviation among the residuals than from other sites (0.025 degrees).

Finally, the frame rate on this camera changed after the final two flares. The cumulative timing post-flare is therefore uncertain, with the resulting total fireball duration being at least 4.5 sec. Speeds and timing agree with other cameras prior to the flares as well as with the final GLM flare timing. Because of this uncertainty we omit any velocity or timing comparisons post-flare in our analysis as these are all based on Station 95, but provide our best estimate of the timing for station 95 from available information in Table A1.

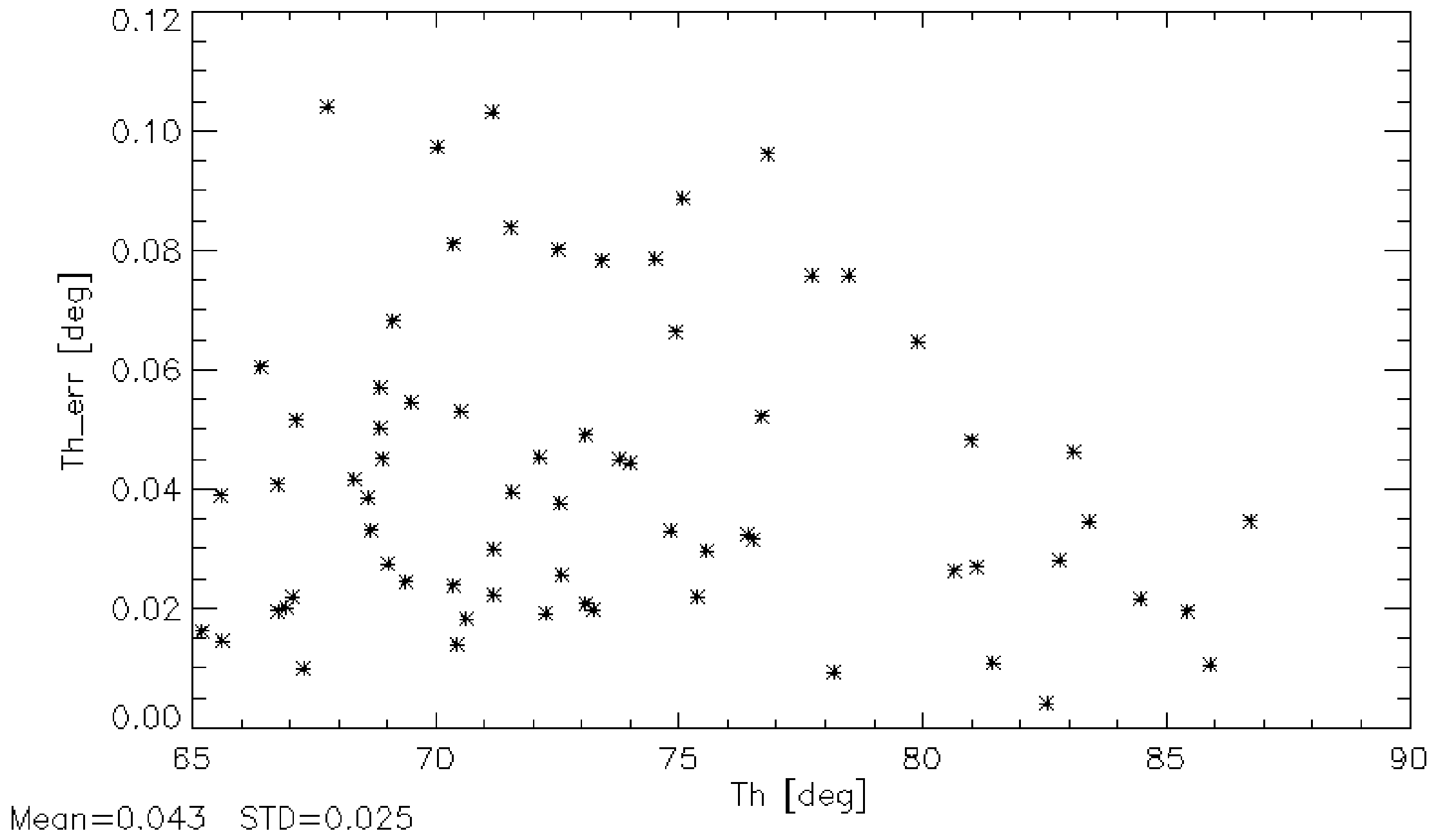

Figure A7. Fit residuals between the plate and individual stars used for calibration for station 95 as a function of zenith distance.

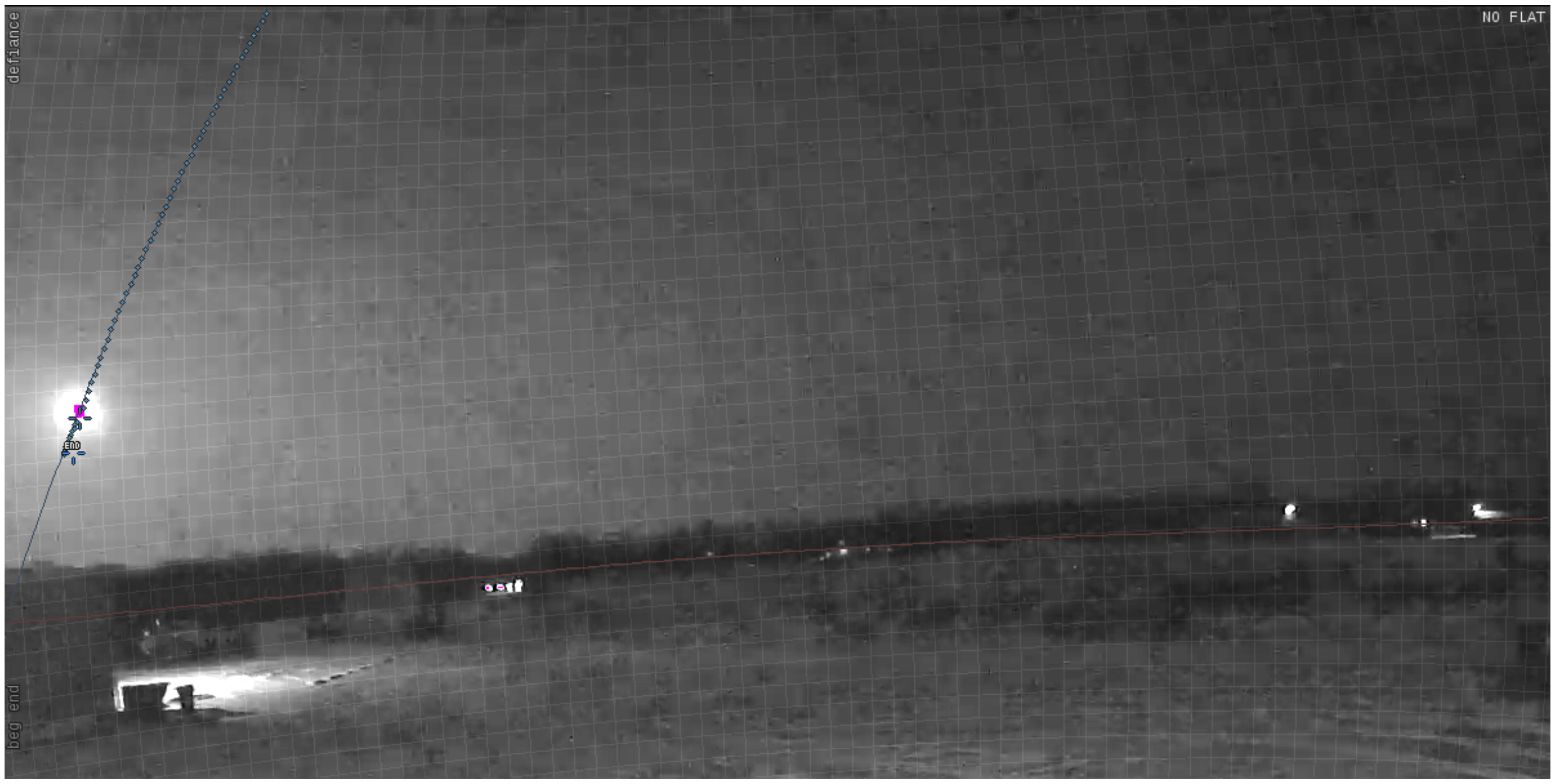

Fig A8. Plate fit for Defiance, OH [95] showing altitude/azimuth gird lines in one degree increments. Also shown is the plate projected final trajectory great circle for the fireball (blue line), its begin and end points and the true horizon (orange line).

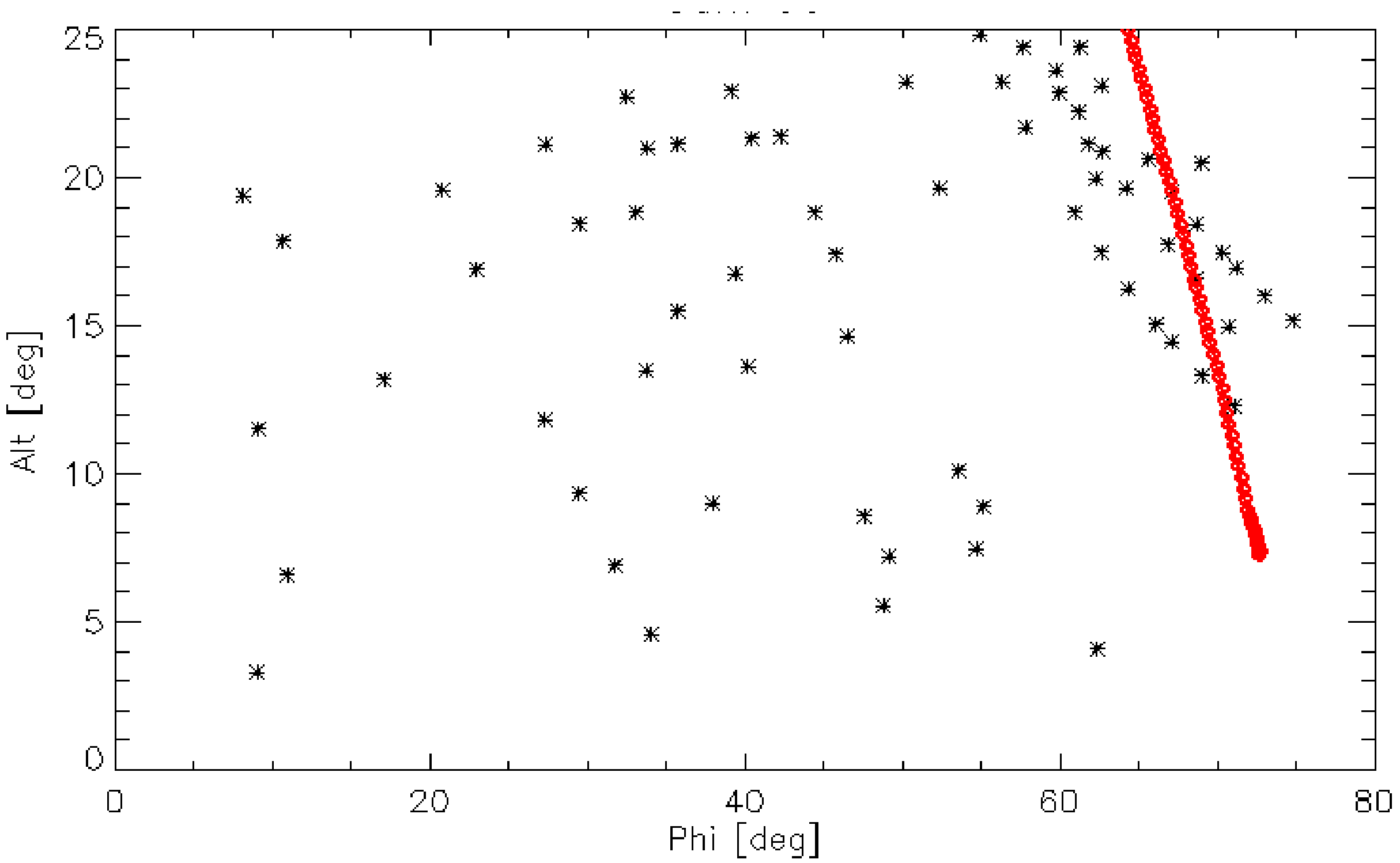

Figure A9. Local coordinates of star picks (black dots) relative to the apparent fireball picks (red dots) for Defiance, OH.

<u>Station 99– Brant, Ontario, Canada</u>

The private security video from this site captured the fireball near the center of the camera field of view. Calibration data from the night of Jan 25, 2018 between 7-11 UT was used to measure 142 stellar positions. As shown in Figure A10, the average fit residuals is 0.03 degrees. The last two frames showing the fireball are heavily saturated and reliable picks are not possible, but the first four frames allow reasonable centroid picks. Figure A11 shows the overall plate which is quite good, reflecting a good distribution of calibration stars around the fireball path (Figure A12). Note that the left hand portion of the video has no calibration stars and the plate is unreliable here (shown by the extension of the apparent horizon upward into open sky).

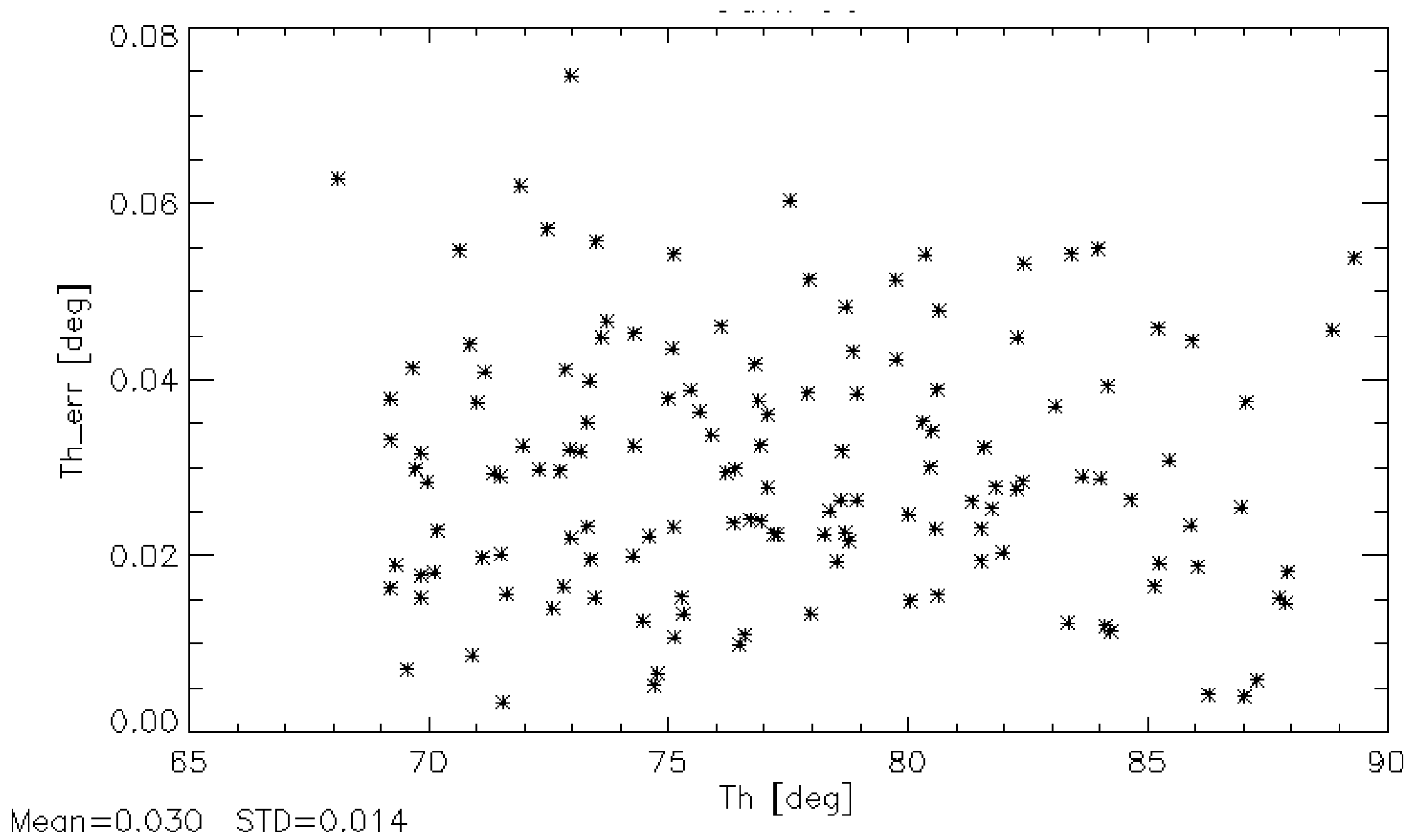

Figure A10. Fit residuals between the plate and individual stars used for calibration for station 99 as a function of zenith distance.

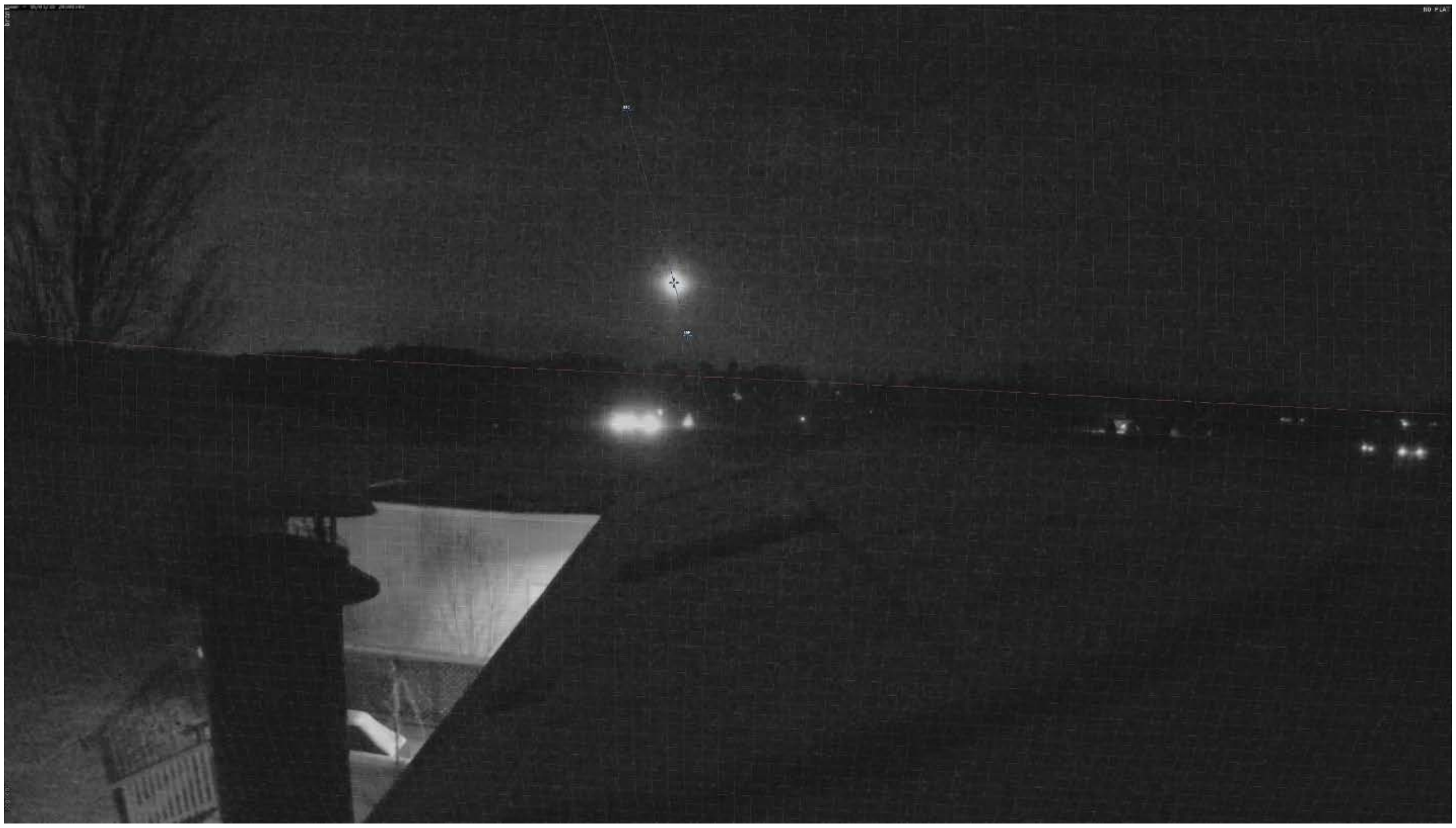

Figure A11. Plate fit for Brant, ON [99] showing altitude/azimuth gird lines in one degree increments. Also shown is the final trajectory plane for the fireball (blue line), its begin and end points and the true horizon (orange line).

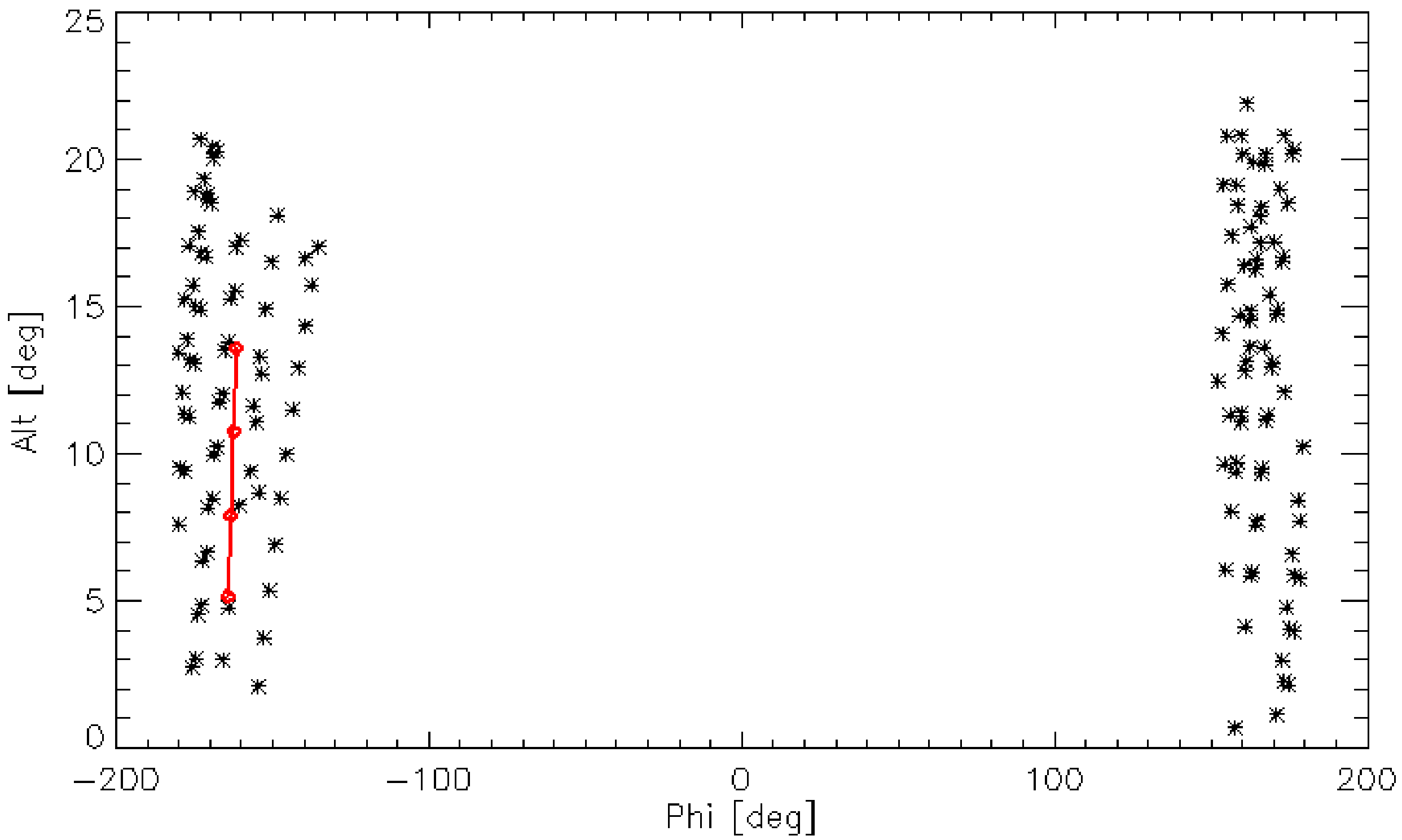

Fig A12. Local coordinates of star picks (black dots) relative to the apparent fireball picks (red dots) for Brant, ON.

Trajectory Solution

Using the individual astrometric measurements from each station (Table A1) a best fit trajectory was found using the least-squares line of sight method of Borovicka (1990). The implementation of this method is described in detail in Vida et al (2019). The best fit trajectory is simply the 3D line where the residuals from all sight lines are minimized. The horizontal and vertical residuals of all sightlines from this best fit trajectory are given in Figure A12. The resulting ground trajectory and local radiant are presented in the main text in Table 2. Uncertainties were found by examining the change in radiant and speed when removing outlying points from some stations (such as the first ~30 points from Chicago) and comparing to the baseline solution. A lower bound to the uncertainty in the orbit/trajectory was determined by generating 1000 Monte Carlo clones using the original sight line measurements and adding Gaussian noise to each measurement with a spread corresponding to the standard deviation of the average sightline residual per station. The corresponding spread in all uncertainties are found from this Monte Carlo ensemble and represent the error in the mathematical fit to the trajectory, providing a lower bound to the overall error. The associated spread in geocentric radiants is shown in Figure A13; the corresponding orbital covariance matrix is given in Table A2. Finally, we use the standard deviation in the individual speeds measured per station as a least squares fit to the length vs. time from the best fit trajectory solution above 50 km altitude as shown in Table A3 to estimate an upper bound to the uncertainty in speed.

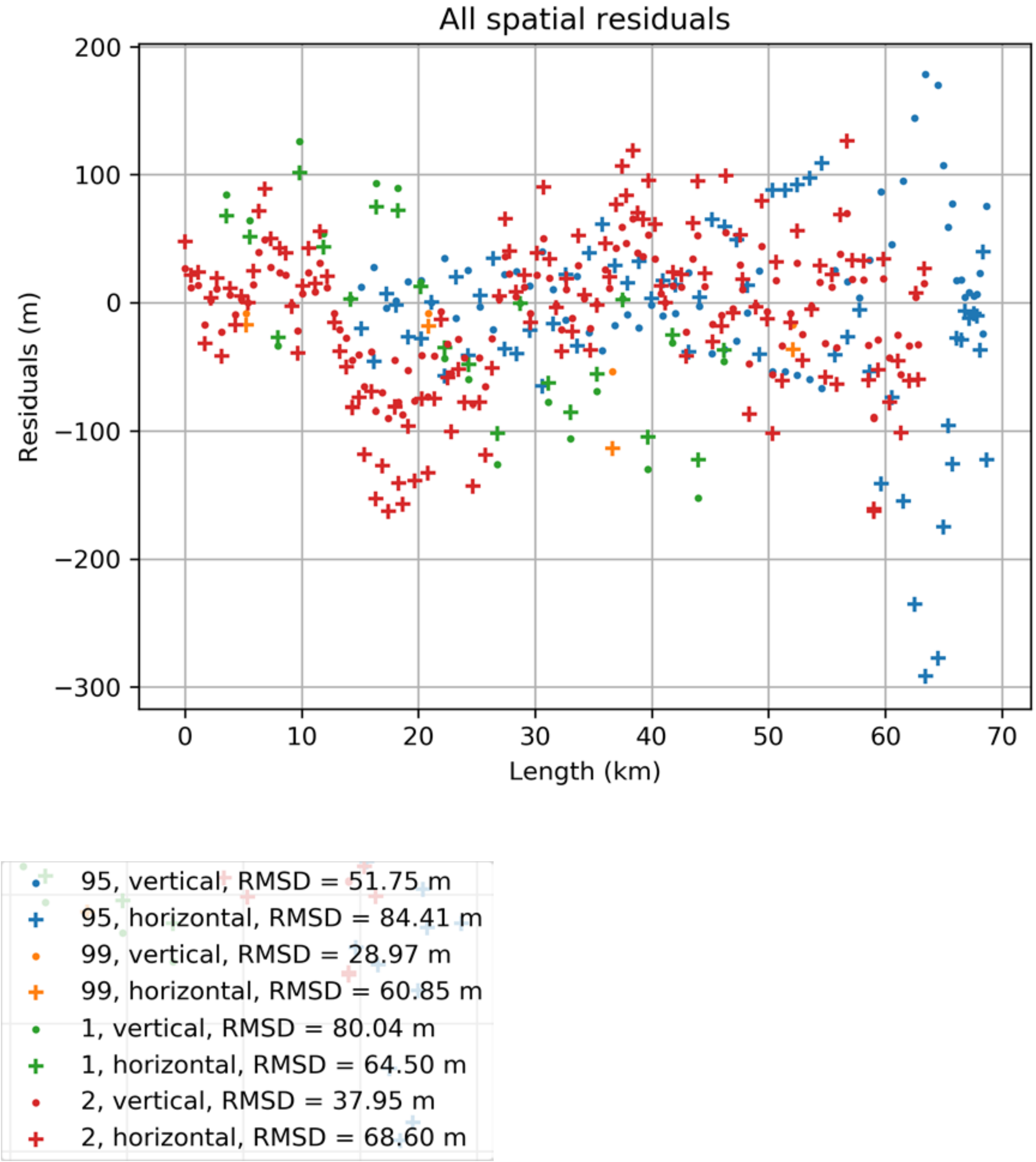

Figure A12. Horizontal and vertical residuals as a function of path length for the best linear-least squares fit to all measured sight lines from four stations. The legend includes the root-mean squared deviation for both vertical and horizontal residuals for all four stations.

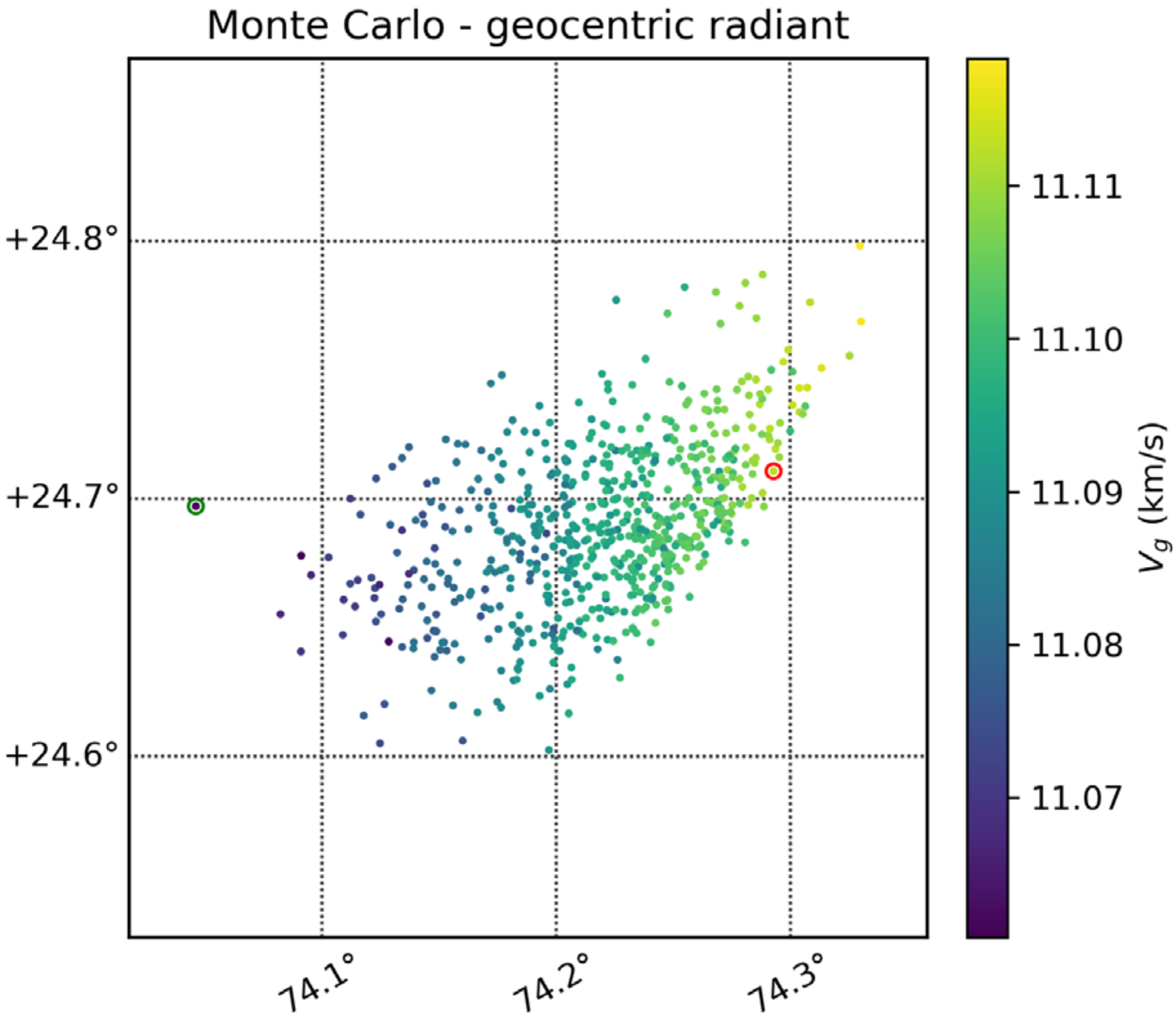

Figure A13. Spread in geocentric radiants for 1000 Monte Carlo clones of the Hamburg fireball. The best fit geometrical radiant is shown with a red circle. The green dot represents the best fit radiant solution weighted by apparent trajectory lag (not used in the final solution, but shown here for comparison).

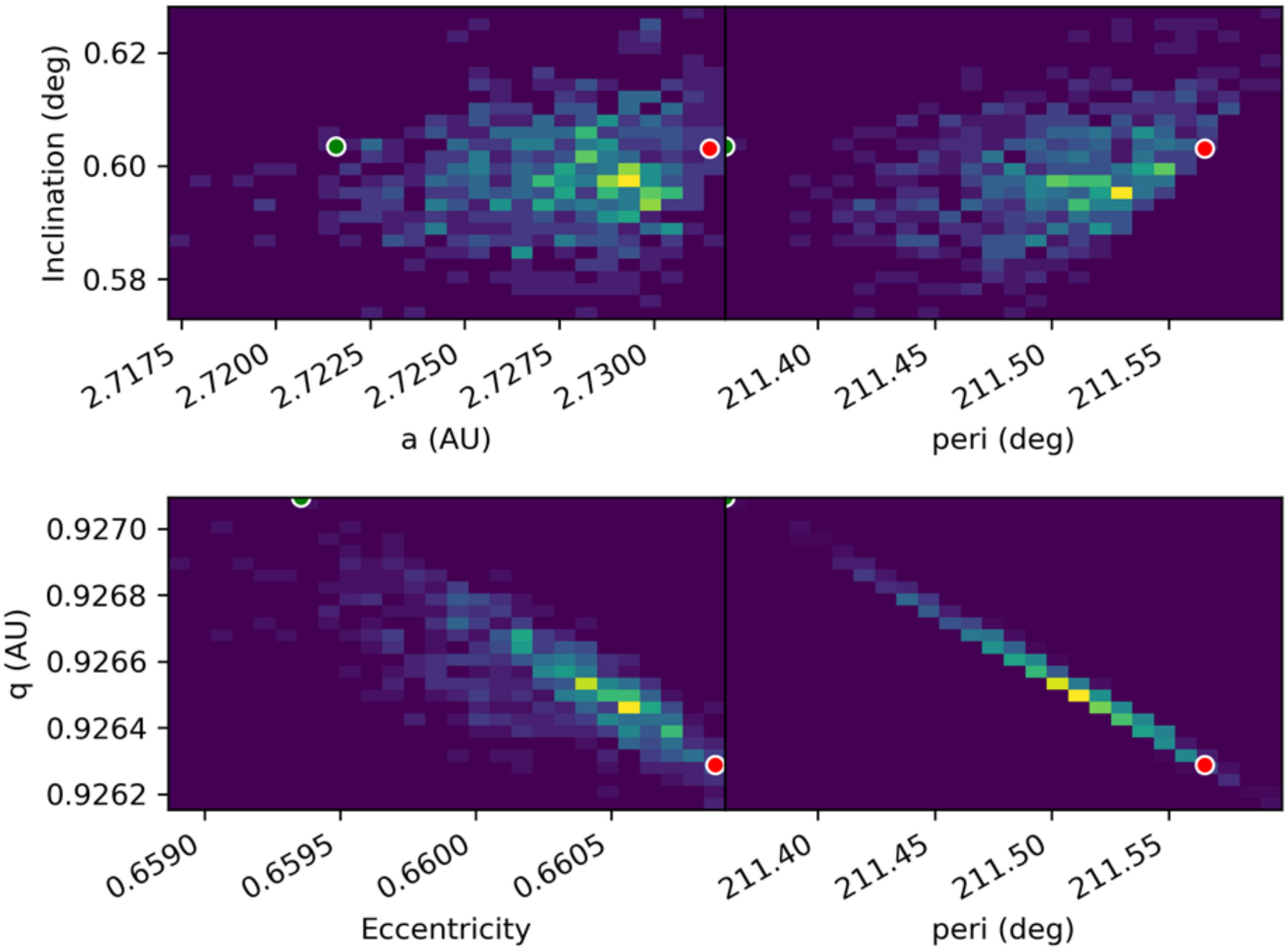

Figure A14. Orbital element dependencies for the uncertainty in the Hamburg orbit, following the covariance matrix in Table A2. The red circle here represents the best fit geometric solution, while the green dot represents the best fit weighted by apparent trajectory lag (not used in the final solution, but shown here for comparison)

Table A1. Individual altitude and azimuth picks for the Hamburg fireball for each station with relative times shown, including timing offsets.

| Point No | Station ID | Time (s) | Azim +E of due N (deg) | Alt (deg) |
|---|---|---|---|---|
| 0 | 95 | 0.937884 | 26.168 | 25.916 |
| 1 | 95 | 1.004584 | 25.997 | 25.573 |
| 2 | 95 | 1.071384 | 25.793 | 25.245 |
| 3 | 95 | 1.138084 | 25.652 | 24.975 |
| 4 | 95 | 1.204784 | 25.49 | 24.645 |
| 5 | 95 | 1.271584 | 25.308 | 24.298 |
| 6 | 95 | 1.338284 | 25.141 | 24.011 |
| 7 | 95 | 1.404984 | 24.981 | 23.644 |
| 8 | 95 | 1.471784 | 24.781 | 23.339 |
| 9 | 95 | 1.538484 | 24.641 | 23.005 |
| 10 | 95 | 1.605184 | 24.446 | 22.674 |
| 11 | 95 | 1.671984 | 24.244 | 22.307 |
| 12 | 95 | 1.738684 | 24.109 | 21.967 |
| 13 | 95 | 1.805384 | 23.941 | 21.631 |
| 14 | 95 | 1.872184 | 23.744 | 21.256 |
| 15 | 95 | 1.938884 | 23.581 | 20.882 |
| 16 | 95 | 2.005684 | 23.405 | 20.582 |
| 17 | 95 | 2.072384 | 23.205 | 20.218 |
| 18 | 95 | 2.139084 | 23.066 | 19.876 |
| 19 | 95 | 2.205884 | 22.861 | 19.536 |
| 20 | 95 | 2.272584 | 22.667 | 19.161 |
| 21 | 95 | 2.339284 | 22.505 | 18.791 |
| 22 | 95 | 2.406084 | 22.331 | 18.414 |
| 23 | 95 | 2.472784 | 22.157 | 18.071 |
| 24 | 95 | 2.539484 | 21.993 | 17.695 |
| 25 | 95 | 2.606284 | 21.821 | 17.35 |
| 26 | 95 | 2.672984 | 21.644 | 16.972 |
| 27 | 95 | 2.739684 | 21.479 | 16.561 |
| 28 | 95 | 2.806484 | 21.31 | 16.252 |
| 29 | 95 | 2.873184 | 21.104 | 15.879 |
| 30 | 95 | 2.939884 | 20.928 | 15.494 |
| 31 | 95 | 3.006684 | 20.759 | 15.118 |
| 32 | 95 | 3.073384 | 20.619 | 14.776 |
| 33 | 95 | 3.140084 | 20.475 | 14.402 |
| 34 | 95 | 3.206884 | 20.229 | 14.011 |
| 35 | 95 | 3.273584 | 20.055 | 13.63 |
| 36 | 95 | 3.340284 | 19.88 | 13.25 |
| 37 | 95 | 3.407084 | 19.704 | 12.868 |
| 38 | 95 | 3.473784 | 19.527 | 12.49 |
| 39 | 95 | 3.540484 | 19.41 | 12.054 |
| 40 | 95 | 3.607284 | 19.231 | 11.672 |
| 41 | 95 | 3.673984 | 19.053 | 11.299 |
| 42 | 95 | 3.740684 | 18.929 | 10.968 |
| 43 | 95 | 3.807484 | 18.806 | 10.591 |
| 44 | 95 | 3.874184 | 18.625 | 10.265 |
| 45 | 95 | 3.940884 | 18.5 | 9.889 |
| 46 | 95 | 4.007684 | 18.373 | 9.508 |
| 47 | 95 | 4.074384 | 18.243 | 9.149 |
| 48 | 95 | 4.141084 | 18.063 | 8.756 |
| 49 | 95 | 4.174484 | 17.941 | 8.601 |
| 50 | 95 | 4.207884 | 17.842 | 8.472 |
| 51 | 95 | 4.241284 | 17.791 | 8.323 |
| 52 | 95 | 4.274584 | 17.69 | 8.212 |
| 53 | 95 | 4.307984 | 17.621 | 8.055 |
| 54 | 95 | 4.341384 | 17.561 | 7.949 |
| 55 | 95 | 4.374684 | 17.502 | 7.81 |
| 56 | 95 | 4.408084 | 17.433 | 7.659 |
| 57 | 95 | 4.441484 | 17.389 | 7.561 |
| 58 | 95 | 4.474784 | 17.359 | 7.466 |
| 59 | 95 | 4.508184 | 17.28 | 7.382 |
| 60 | 95 | 4.541584 | 17.303 | 7.246 |
| 0 | 99 | 0.323157 | 251.513 | 13.565 |
| 1 | 99 | 1.323157 | 252.481 | 10.746 |
| 2 | 99 | 2.323157 | 253.416 | 7.906 |
| 3 | 99 | 3.323157 | 254.345 | 5.159 |
| 0 | 1 | 0 | 97.824 | 7.087 |
| 1 | 1 | 0.1335 | 97.795 | 6.895 |
| 2 | 1 | 0.2669 | 97.77 | 6.657 |
| 3 | 1 | 0.4004 | 97.717 | 6.485 |
| 4 | 1 | 0.5339 | 97.694 | 6.281 |
| 5 | 1 | 0.6673 | 97.664 | 6.055 |
| 6 | 1 | 0.8008 | 97.614 | 5.838 |
| 7 | 1 | 0.9343 | 97.584 | 5.654 |
| 8 | 1 | 1.0677 | 97.563 | 5.459 |
| 9 | 1 | 1.2012 | 97.538 | 5.253 |
| 10 | 1 | 1.3347 | 97.506 | 5.045 |
| 11 | 1 | 1.4681 | 97.475 | 4.797 |
| 12 | 1 | 1.6016 | 97.423 | 4.601 |
| 13 | 1 | 1.7351 | 97.394 | 4.354 |
| 14 | 1 | 1.8685 | 97.366 | 4.158 |
| 15 | 1 | 2.002 | 97.322 | 3.927 |
| 16 | 1 | 2.1355 | 97.273 | 3.7 |
| 17 | 1 | 2.2689 | 97.256 | 3.471 |
| 18 | 1 | 2.4024 | 97.205 | 3.256 |
| 19 | 1 | 2.5359 | 97.185 | 3.02 |
| 20 | 1 | 2.6693 | 97.131 | 2.795 |
| 0 | 2 | 0 | 80.006 | 12.048 |
| 1 | 2 | 0.0334 | 79.988 | 11.976 |
| 2 | 2 | 0.0667 | 79.963 | 11.9 |
| 3 | 2 | 0.1001 | 79.949 | 11.822 |
| 4 | 2 | 0.1335 | 79.921 | 11.757 |
| 5 | 2 | 0.1668 | 79.896 | 11.689 |
| 6 | 2 | 0.2002 | 79.89 | 11.633 |
| 7 | 2 | 0.2336 | 79.849 | 11.538 |
| 8 | 2 | 0.2669 | 79.835 | 11.477 |
| 9 | 2 | 0.3003 | 79.808 | 11.407 |
| 10 | 2 | 0.3337 | 79.785 | 11.332 |
| 11 | 2 | 0.367 | 79.761 | 11.273 |
| 12 | 2 | 0.4004 | 79.733 | 11.215 |
| 13 | 2 | 0.4338 | 79.707 | 11.145 |
| 14 | 2 | 0.4671 | 79.692 | 11.074 |
| 15 | 2 | 0.5005 | 79.661 | 10.972 |
| 16 | 2 | 0.5339 | 79.639 | 10.901 |
| 17 | 2 | 0.5672 | 79.624 | 10.828 |
| 18 | 2 | 0.6006 | 79.608 | 10.755 |
| 19 | 2 | 0.634 | 79.582 | 10.707 |
| 20 | 2 | 0.6673 | 79.554 | 10.638 |
| 21 | 2 | 0.7007 | 79.534 | 10.558 |
| 22 | 2 | 0.7341 | 79.51 | 10.509 |
| 23 | 2 | 0.7674 | 79.49 | 10.424 |
| 24 | 2 | 0.8008 | 79.468 | 10.332 |
| 25 | 2 | 0.8342 | 79.454 | 10.274 |
| 26 | 2 | 0.8675 | 79.431 | 10.194 |
| 27 | 2 | 0.9009 | 79.415 | 10.124 |
| 28 | 2 | 0.9343 | 79.389 | 10.047 |
| 29 | 2 | 0.9676 | 79.375 | 9.975 |
| 30 | 2 | 1.001 | 79.341 | 9.899 |
| 31 | 2 | 1.0344 | 79.339 | 9.84 |
| 32 | 2 | 1.0677 | 79.31 | 9.765 |
| 33 | 2 | 1.1011 | 79.294 | 9.692 |
| 34 | 2 | 1.1345 | 79.255 | 9.621 |
| 35 | 2 | 1.1678 | 79.251 | 9.571 |
| 36 | 2 | 1.2012 | 79.239 | 9.523 |
| 37 | 2 | 1.2346 | 79.206 | 9.458 |
| 38 | 2 | 1.2679 | 79.189 | 9.377 |
| 39 | 2 | 1.3013 | 79.152 | 9.301 |
| 40 | 2 | 1.3347 | 79.14 | 9.226 |
| 41 | 2 | 1.368 | 79.103 | 9.146 |
| 42 | 2 | 1.4014 | 79.067 | 9.072 |
| 43 | 2 | 1.4348 | 79.051 | 8.992 |
| 44 | 2 | 1.4681 | 79.045 | 8.947 |
| 45 | 2 | 1.5015 | 79.009 | 8.864 |
| 46 | 2 | 1.5349 | 78.989 | 8.784 |
| 47 | 2 | 1.5682 | 78.97 | 8.681 |
| 48 | 2 | 1.6016 | 78.933 | 8.606 |
| 49 | 2 | 1.635 | 78.918 | 8.532 |
| 50 | 2 | 1.6683 | 78.879 | 8.452 |
| 51 | 2 | 1.7017 | 78.842 | 8.371 |
| 52 | 2 | 1.7351 | 78.807 | 8.299 |
| 53 | 2 | 1.7684 | 78.795 | 8.245 |
| 54 | 2 | 1.8018 | 78.777 | 8.167 |
| 55 | 2 | 1.8352 | 78.742 | 8.064 |
| 56 | 2 | 1.8685 | 78.726 | 7.989 |
| 57 | 2 | 1.9019 | 78.69 | 7.91 |
| 58 | 2 | 1.9353 | 78.654 | 7.829 |
| 59 | 2 | 1.9686 | 78.641 | 7.751 |
| 60 | 2 | 2.002 | 78.624 | 7.672 |
| 61 | 2 | 2.0354 | 78.609 | 7.602 |
| 62 | 2 | 2.0687 | 78.58 | 7.547 |
| 63 | 2 | 2.1021 | 78.564 | 7.469 |
| 64 | 2 | 2.1355 | 78.526 | 7.397 |
| 65 | 2 | 2.1688 | 78.512 | 7.322 |
| 66 | 2 | 2.2022 | 78.497 | 7.246 |
| 67 | 2 | 2.2356 | 78.465 | 7.167 |
| 68 | 2 | 2.2689 | 78.422 | 7.061 |
| 69 | 2 | 2.3023 | 78.411 | 7.009 |
| 70 | 2 | 2.3357 | 78.374 | 6.928 |
| 71 | 2 | 2.369 | 78.344 | 6.852 |
| 72 | 2 | 2.4024 | 78.332 | 6.799 |
| 73 | 2 | 2.4358 | 78.301 | 6.723 |
| 74 | 2 | 2.4691 | 78.291 | 6.659 |
| 75 | 2 | 2.5025 | 78.271 | 6.592 |
| 76 | 2 | 2.5359 | 78.244 | 6.526 |
| 77 | 2 | 2.5692 | 78.223 | 6.436 |
| 78 | 2 | 2.6026 | 78.209 | 6.359 |
| 79 | 2 | 2.636 | 78.195 | 6.306 |
| 80 | 2 | 2.6693 | 78.159 | 6.206 |
| 81 | 2 | 2.7027 | 78.127 | 6.102 |
| 82 | 2 | 2.7361 | 78.119 | 6.034 |
| 83 | 2 | 2.7694 | 78.074 | 5.959 |
| 84 | 2 | 2.8028 | 78.049 | 5.901 |
| 85 | 2 | 2.8362 | 78.032 | 5.798 |
| 86 | 2 | 2.8695 | 78.013 | 5.701 |
| 87 | 2 | 2.9029 | 77.973 | 5.58 |
| 88 | 2 | 2.9363 | 77.934 | 5.534 |
| 89 | 2 | 2.9696 | 77.926 | 5.436 |
| 90 | 2 | 3.003 | 77.885 | 5.345 |
| 91 | 2 | 3.0364 | 77.882 | 5.313 |
| 92 | 2 | 3.0697 | 77.876 | 5.223 |
| 93 | 2 | 3.1031 | 77.835 | 5.147 |
| 94 | 2 | 3.1365 | 77.794 | 5.071 |
| 95 | 2 | 3.1698 | 77.789 | 4.993 |
| 96 | 2 | 3.2032 | 77.784 | 4.917 |
| 97 | 2 | 3.2366 | 77.745 | 4.882 |
| 98 | 2 | 3.2699 | 77.738 | 4.797 |
| 99 | 2 | 3.3033 | 77.696 | 4.692 |
| 100 | 2 | 3.3367 | 77.655 | 4.608 |
| 101 | 2 | 3.37 | 77.652 | 4.53 |
| 102 | 2 | 3.4034 | 77.609 | 4.418 |
| 103 | 2 | 3.4368 | 77.567 | 4.305 |
| 104 | 2 | 3.4701 | 77.562 | 4.23 |
| 105 | 2 | 3.5035 | 77.521 | 4.152 |
| 106 | 2 | 3.5369 | 77.516 | 4.078 |
| 107 | 2 | 3.5702 | 77.477 | 4.043 |
| 108 | 2 | 3.6036 | 77.438 | 3.956 |
| 109 | 2 | 3.637 | 77.435 | 3.883 |
| 110 | 2 | 3.6703 | 77.389 | 3.733 |
| 111 | 2 | 3.7037 | 77.386 | 3.66 |
| 112 | 2 | 3.7371 | 77.383 | 3.582 |
| 113 | 2 | 3.7704 | 77.383 | 3.582 |
| 114 | 2 | 3.8038 | 77.345 | 3.535 |
| 115 | 2 | 3.8372 | 77.305 | 3.465 |
| 116 | 2 | 3.8705 | 77.303 | 3.382 |
| 117 | 2 | 3.9039 | 77.262 | 3.271 |
| 118 | 2 | 3.9373 | 77.26 | 3.227 |
| 119 | 2 | 3.9706 | 77.217 | 3.115 |
| 120 | 2 | 4.004 | 77.178 | 3.036 |
| 121 | 2 | 4.0374 | 77.18 | 2.997 |
| 122 | 2 | 4.0707 | 77.139 | 2.924 |

Table A2. Orbital covariance matrix for best-fit trajectory solution for the Hamburg fireball. Units of Tp are in days.

| | e | q (AU) | Tp (JD) | node (rad) | peri (rad) | i (rad) |
|---|---|---|---|---|---|---|
| e | 1.27E-07 | -4.26E-08 | -8.03E-04 | 2.18E-09 | 1.75E-07 | 1.22E-08 |
| q | -4.26E-08 | 2.33E-08 | 2.44E-04 | -1.91E-09 | -1.01E-07 | -1.06E-08 |
| Tp | -8.03E-04 | 2.44E-04 | 5.15E+00 | -1.04E-05 | -9.81E-04 | -5.83E-05 |
| node | 2.18E-09 | -1.91E-09 | -1.04E-05 | 7.55E-10 | 8.07E-09 | 4.19E-09 |
| peri | 1.75E-07 | -1.01E-07 | -9.81E-04 | 8.07E-09 | 4.45E-07 | 4.48E-08 |
| i | 1.22E-08 | -1.06E-08 | -5.83E-05 | 4.19E-09 | 4.48E-08 | 2.33E-08 |

Table A3. Average speeds for the fireball for each of the four stations above ~50 km height. The Velocity (LS) uses a least squares fit to the length vs time for each point from each station above the measurement height in an ECF frame (relative to the fixed Earth).

| Station | Beg Height (km) | Measurement Height (km) | Velocity LS (km/s) |
|---|---|---|---|
| 95 | 69.1 | 50.1 | 15.58 |
| 99 | 78.2 | 49.3 | 15.76 |
| 1 | 80.7 | 50.5 | 15.9 |
| 2 | 83.0 | 49.8 | 15.86 |
| | | Average | 15.78 |
| | | Standard Deviation | 0.14 |
| | | | |